\documentclass[fleqn,usenatbib,useAMS]{mnras}

\usepackage{graphicx}	
\usepackage{amsmath}	
\usepackage{amssymb}	
\usepackage{multicol}        
\usepackage{bm}		
\usepackage{pdflscape}	
\usepackage{subcaption}

\newcommand{\stokes}{{\mathrm{St}}}

\usepackage[T1]{fontenc}
\usepackage{ae,aecompl}

\usepackage{newtxtext,newtxmath}

\title[Dusty Ring Stability]{Stability of Dusty Rings in Protoplanetary Discs}

\author[K . Chan and S-J . Paardekooper]{Kevin Chan$^{1}$\thanks{Contact e-mail: \href{mailto:k.chan@qmul.ac.uk}{k.chan@qmul.ac.uk}} and Sijme-Jan Paardekooper$^{2,1}$\\
$^{1}$Astronomy Unit, School of Physical and Chemical Sciences, Queen Mary University of London, Bethnal Green, London E1 4NS\\
$^{2}$Faculty of Aerospace Engineering, Delft University of Technology, Kluyverweg 1, 2629 HS Delft, The Netherlands}

\date{Accepted 2024 January 4. Received 2023 December 1; in original form 2023 June 4}

\pubyear{2024}

\begin{document}
\label{firstpage}
\pagerange{\pageref{firstpage}--\pageref{lastpage}}
\maketitle

\begin{abstract}
Dust rings in protoplanetary discs are often observed in thermal dust emission and could be favourable environments for planet formation. While dust rings readily form in gas pressure maxima, their long-term stability is key to both their observability and potential to assist in planet formation. We investigate the stability of the dust ring generated by interactions of a protoplanetary disc with a Neptune sized planet and consider its possible long term evolution using the FARGO3D Multifluid code. We look at the onset of the Rossby Wave Instability (RWI) and compare how the addition of dust in a disc can alter the stability of the gas phase. We find that with the addition of dust, the rings generated by planet disc interactions are more prone to RWI and can cause the gas phase to become unstable. The instability is shown to occur more easily for higher Stokes number dust, as it accumulates into a more narrow ring which triggers the RWI, while the initial dust fraction plays a more minor role in the stability properties. We show that the dusty RWI generates vortices that collect dust in their cores, which could be sites for further planetesimal formation. We conclude that the addition of dust can cause a ring in a protoplanetary disc to become more prone to instability leading to a different long term evolution compared to gas only simulations of the RWI.

\end{abstract}

\begin{keywords}
protoplanetary discs - hydrodynamics - method: numerical - planet-disc interactions
\end{keywords}



\section{Introduction} \label{intro}

Radial drift of dust particles in protoplanetary discs (e.g., \citealt{weidenschilling77}) is a key process during planet formation. This radial motion is driven by aerodynamic drag with the gaseous component of the disc, and it transports dust particles to regions of high gas pressure, which is usually towards the central star. The drift speed depends on particle size, with meter sized boulders typically reaching the highest speeds, for which drift time scales can be as short as 100 years at 1 AU. As the dust particles drift inwards, they can gather and clump together, with the larger particles travelling at a higher drift speed. This can lead to fragmentation where collisions occur with other particles, but, very importantly, the larger particles can drift into the star on a short timescale preventing further growth (\citealt{weidenschilling77}). This generates a problem called the `meter barrier' and inhibits planet formation. A way to overcome the barrier is the presence of pressure bumps and formation of dusty (gas and dust) rings. This provides an environment in the disc which is more favourable towards the formation of planetesimals and growth towards planetary cores (see e.g. \citealt{lyra09}). 

The formation of dust rings and gaps in protoplanetary discs has been widely debated, in particular since observations of the millimeter dust continuum emission from the HL Tau region in \cite{alma15}. These kind of substructures have been found in many discs and characterised in later studies (e.g. \citealt{andrews16}; \citealt{dipierro18}; \citealt{andrews18}; \citealt{huang18}). Observational evidence indicates that mass retention for example is different in structured discs compared to unstructured discs (\citealt{marel21}). For a recent overview, see \cite{bae22}. 

\cite{whipple72} showed that a pressure bump in the gas phase of the disc can act as a trap for dust flowing into the region. The dust grains radially drift towards pressure maxima and can be collected into a ring structure. There are however many possible ways that a pressure bump can exist in a disc. Possible mechanisms include zonal winds (e.g. \citealt{johansen09}; \citealt{flock15}; \citealt{hu22}) whereby large scale variations of the magnetic field through the disc creates regions of slower or faster flows of the rotating gas, interactions with a disc by an embedded planet via its gravitational torque (e.g. \citealt{zhu12}; \citealt{dipierro15}; \citealt{dong17}), frost lines condensation fronts (e.g. \citealt{zhang15}; \citealt{yen16}) as the local surface density near frost lines are enhanced, edges of dead zones where a differential mass accretion at the edges can create a density bump (\citealt{varniere06}) and large scale instabilities (e.g. \citealt{aguilar15}), with the example given being dust settling creating a dynamical dust-gas instability in the disc that leads to toroidal vortices.

Regions of enhanced dust concentration, such as pressure bumps, are naturally favourable sites for planet formation. A locally enhanced dust to gas ratio in the ring can trigger the Streaming Instability (e.g., \citealt{youdin05}; \citealt{johansen07}; \citealt{bai10}) which has been shown to overcome the 'meter-size' barrier by accumulating enough particles for gravitational instability to occur. Additionally, self gravity can cause or aid the growth of planetesimals in the region (\citealt{chiang10}; \citealt{simon16}; \citealt{gerbig20}). The Streaming Instability works as a drag instability with the dust feeling a headwind from the gas as it orbits at sub-Keplerian speeds in a disc (see e.g. \citealt{squire20}). This can cause a traffic jam with dust migrating inwards and clumping together. With larger clumps providing a stronger back-reaction on the gas, this can reduce the local headwind meaning isolated particles further out can drift inwards faster and join the clump causing the same process to occur again. The Streaming Instability is one of the popular mechanisms; however, an alternative mechanism includes the use of vortices to concentrate the dust in the centre (\citealt{barge95}), which have been a highly common feature in hydrodynamical simulations and possible explanation for asymmetric substructures in observations (\citealt{marr22}). The issues with drag instabilities and using vortices as an alternative with aiding planet formation are discussed and explored in \cite{lovascio22}.

In addition to ring-like structures, many protoplanetary discs have shown large scale azimuthal asymmetries in the mm dust continuum (e.g. \citealt{marel13}; \citealt{isella13}; \citealt{perez14}). A possible explanation for these structures is the existence of vortices in the disc, triggered through the Rossby Wave Instability (RWI) which has been studied analytically (\citealt{lovelace99}; \citealt{li00b}) and numerically (\citealt{li00a}; \citealt{varniere06}). This is a mechanism where an axisymmetric bump in the density profile of the disc (\citealt{meheut12a}) and/or steep density gradients (\citealt{koller03}) can give rise to high pressure anticyclonic vortices. These vortices can not only act as dust traps in the centre (\citealt{barge95}) but also smooth out the density extrema which trigger the instability through angular momentum exchange. This has an important consequence as gas rings in the protoplanetary disc may be prone to the RWI if the density gradient is too high, and thus smooth out over time through the action of the RWI, therefore inhibiting planetesimal formation at the location of the ring. This raises the question of how stable are the rings and gaps in protoplanetary discs. So far, studies of the stability of dusty rings (e.g. \citealt{huang20}; \citealt{pierens21}; \citealt{lee22}) have shown that a wide range of instabilities can occur in and around rings in a disc as it evolves over time. Previously, \cite{meheut12a} studied the formation and long term evolution of vortices in a gas only protoplanetary disc however the inclusion of dust in a disc and how it affects how prone the ring may be to RWI has not been studied well. Additionally in \cite{cimerman23}, they study the stability of gaps in a gas only protoplanetary disc where vortices have been shown to emerge at the edge of planet driven gaps.

For simulations which include dust and gas, \cite{pierens19} investigated the interaction between a low mass planet and pebble rich disc. They find that dusty vortices can develop through the RWI. Since the RWI can develop at entropy extrema of a gaseous disc, they expected the RWI to be triggered at high dust to gas ratios, which a dusty ring would be an ideal region for. In \cite{huang20}, they run two fluid simulations where they find a meso-scale instability that can break up the dusty ring formed. This instability was caused by a steep velocity shear between the gas inside and outside the ring, as a high dust to gas ratio "forces" the gas in the ring to rotate near Keplerian velocity. Lastly in \cite{hsieh20}, they investigate the evolution of the migration of low mass planets in an inviscid dusty disc. In their study they find that high metallicty discs with large Stokes numbers eventually generates small scale dusty vortices at the planet's gap edges which can halt or reverse the planet's initial inward migration. These studies show that vortices generated can alter the evolution of dusty regions in many different ways. The dust can have a large impact on the size and location of the possible instabilities generated around a planet and in the disc morphology created by an embedded planet.

Clearly, there is an intricate relation between gas pressure bumps, dust rings and vortices, relevant to both observations and theory. \cite{chang23} provided stability calculations of gas pressure bumps, and showed under which conditions pressure bumps can be both stable to the RWI and able to trap dust. They showed that under a wide range of conditions, dust traps are RWI stable and therefore potentially long lived. Most recently in \cite{liu23}, they conducted simulations of dust and gas in protoplanetary discs and found two types of $H$-scale instabilities ($H$ being pressure scale height) where they are termed as the dusty Rossby Wave Instability (DRWI). These were found in pressure bumps where Type I instability dominates relatively sharp pressure bumps and/or bumps with low dust content and Type II, where in a more dusty bump, the DRWI develops anticyclonic vortices smaller than a pressure scale height, largely preserving the dusty ring. In both of these types, the non-linear evolution generates dusty vortices with significant dust mass loading.

In this paper we investigate how the inclusion of dust, and in particular the feedback on the gas, in a protoplanetary disc affects the stability of pressure bumps created by embedded planets and the onset of the Rossby Wave Instability. While for the canonical dust to gas ratio of $1/100$, the dynamics is by and large governed by the gas only, the presence of a dust trap leads to an ever-increasing dust density as long as the trap is fed from the outer disc. At some point, the dust will become dynamically important, which may affect the stability properties of the trap. This is what we would like to investigate in this paper.

The structure of this paper is as follows. In section \ref{method} we detail the simulation setup for the two fluid model in FARGO3D Multifluid. In section \ref{results} we present the results of the effect of the inclusion of dust in a ring generated by a planet and its possible long term evolution. In section \ref{discuss} we discuss our results and the possible implications and conclude in section \ref{conc}.

\section{Method} \label{method}

\subsection{Computational Setup}

We perform gas and dust simulations of a global disc using FARGO3D Multifluid (\citealt{llambay19}). The locally isothermal terminal velocity approximation where the two fluids are combined as a single fluid presented caveats (\citealt{chan22}) when a larger planet was embedded in the disc, therefore we will use the two fluid system where the gas and dust are evolved separately for our simulations. We assume a geometrically thin disc in 2D cylindrical coordinates $(r, \phi)$ with vertically integrated quantities. The equations we solve are the continuity equations for the gas and dust respectively,
\begin{equation} \label{gascont}
    \frac{\partial \Sigma_{\rm g}}{\partial t} + \nabla \cdot (\Sigma_{\rm g} \textbf{u}_{\rm g}) = 0,
\end{equation}

\begin{equation}    \frac{\partial \Sigma_{\rm d}}{\partial t} + \nabla \cdot (\Sigma_{\rm d} \textbf{u}_{\rm d}) = 0,
\end{equation}

\noindent where $\Sigma$ denotes the surface density and $\textbf{u}$ the 2D velocity, with subscripts ${\rm g}$ and ${\rm d}$  referring to the gas and dust. The momentum equations are solved,

\begin{equation}
    \frac{\partial (\Sigma_{\rm g} \textbf{u}_{\rm g})}{\partial t} + \nabla \cdot (\Sigma_{\rm g} \textbf{u}_{\rm g} \textbf{u}_{\rm g}) =  K(\textbf{u}_{\rm d}-\textbf{u}_{\rm g}) - \nabla p_{\rm g} + \Sigma_{\rm g} \nabla\Phi + \nabla\cdot \textbf{T}_{\rm g},
\end{equation}

\begin{equation} \label{dustmom}
    \frac{\partial (\Sigma_{\rm d} \textbf{u}_{\rm d})}{\partial t} + \nabla \cdot (\Sigma_{\rm d} \textbf{u}_{\rm d}\textbf{u}_{\rm d}) =  - K(\textbf{u}_{\rm d}-\textbf{u}_{\rm g}) + \Sigma_{\rm d}\nabla\Phi + \nabla\cdot \textbf{T}_{\rm d}, 
\end{equation}

\noindent with $K$ as the drag coefficient between the two phases and $p_{\rm g}$ the gas pressure. We use a locally isothermal equation of state, $p_{\rm g} = c_s^2(r) \Sigma_{\rm g}$ where $c_s(r)$ is the fixed radial profile of the gas sound speed throughout the disc. Apart from the drag forces, the forces appearing on the right hand side include the gravitational force, expressed through the gravitational potential $\Phi$, and the viscous forces through the stress tensor, $\textbf{T}_i$ which includes $\nu$ as the kinematic viscosity (see Eq. 3 of \citealt{llambay19}) given in units of $r^2_{\rm p} \Omega_k(r_{\rm p})$. Our computational setup does not include dust diffusion and the equations are solved through the method of operator splitting (\citealt{stone92}).

Dimensionless units are adopted whereby the unit of mass is the mass of the primary $M_{\rm P}$. Our standard planet mass corresponds to a Neptune-like planet with mass $M_p=10^{-4} M_{\rm P}$. This corresponds to a ratio of planet to thermal mass of $M_{\rm p}/M_{th} = 0.8$ (\citealt{goodman01}). The radial domain is from 0.4 to 2.5$r_{\rm p}$ (\citealt{devalborro06}) where $r_{\rm p}$ is the orbital radius of the planet. We note that the planet is on a fixed orbit at $1 r_{\rm p}$ to generate the outer pressure bump which we will be analysing in the results. The disc has a constant aspect ratio $h_0=H/r = 0.05$ with $H$ as the scale height and there is no flaring of the disc. The sound speed profile is given as $c_s(r) = h_0\Omega_{\rm K}r$ and the planet potential is smoothed over 0.6$H$. The lowest resolution used in the simulations are (768,1536) cells in the radial and azimuthal direction respectively corresponding to 12 cells per scale height at the planets location. The gas and dust surface density are set to,

\begin{align}
    \Sigma_{\rm g} =& \Sigma_0\Bigg(\frac{r}{r_{\rm p}}\Bigg)^{-0.5}(1-f_{\rm d}),\\
    \Sigma_{\rm d} =& \Sigma_0\Bigg(\frac{r}{r_{\rm p}}\Bigg)^{-0.5}f_{\rm d},
\end{align}

\noindent with $\Sigma_0$ as the total surface density of the gas and dust at the planet's location and $f_{\rm d}$ as the dust fraction. A value of $\Sigma_0 = 6.3662\times10^{-4}$ is used, but this plays no role as self-gravity is not included in these simulations. 

To test the stability of the dust rings we vary the dust fraction and Stokes numbers. The dust species we use in the two fluid simulations are described by a constant Stokes number, $\stokes \in [0.01, 0.05, 0.1, 0.2]$ which characterises the collision rate, $\eta$, between the gas and dust phase given as $\stokes = \Omega_{\rm k}/\eta$ where the collision rate is the inverse of the particle stopping time, $\tau_{\rm s}$ (\citealt{youdin05}). The drag coefficient is related to the collision rate through the equation, $ K = -\Sigma_i \eta$ with subscript $i$ as the $i$-th species, giving us a relation between the Stokes number and drag coefficient as $K = -\Sigma_i \Omega_{\rm k}/\stokes$.

\begin{figure*}
\includegraphics[width=\textwidth]{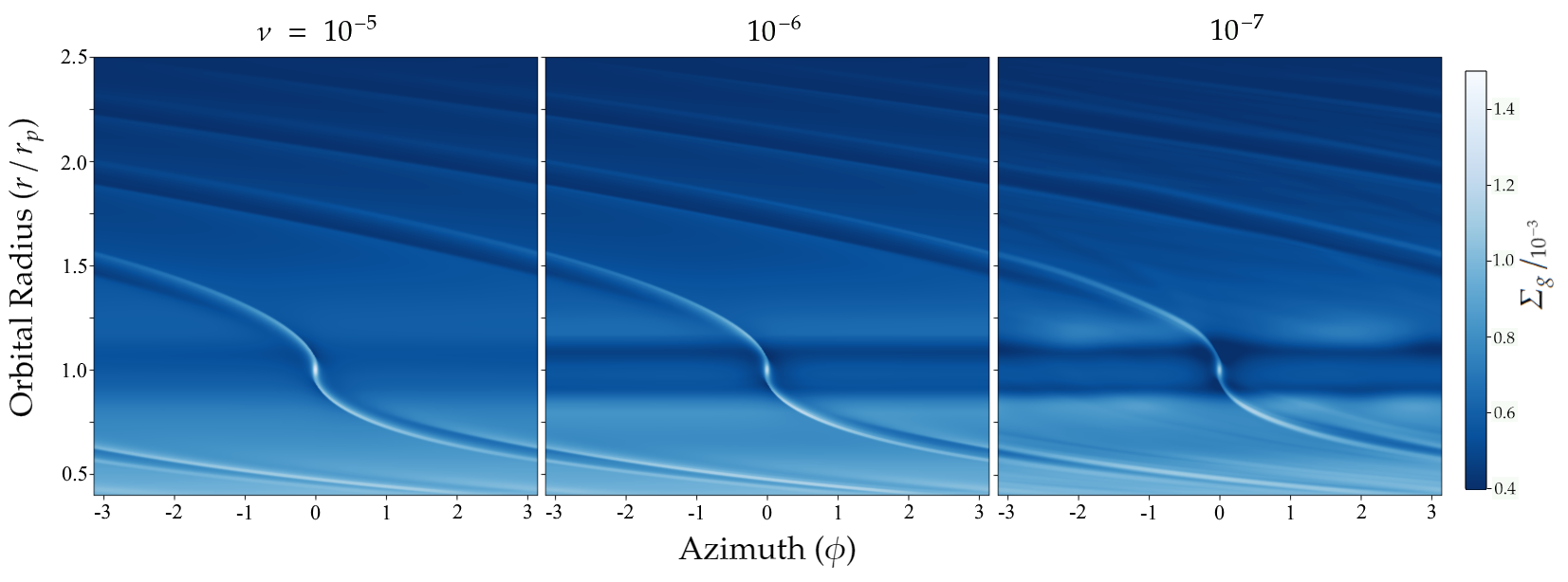}
\caption{Density plots for a gaseous disc with a Neptune sized planet embedded at $1 r_{\rm p}$ after $t = 100$ orbits. The kinematic viscosity is varied, $\nu \in [10^{-5}, 10^{-6}, 10^{-7}]$ from left to right. RWI has developed in the rightmost plot with the lowest kinematic viscosity at the inner and outer rings and has caused vortices to emerge, destabilising the gas rings and gap edges ($r = 0.9, 1.1 r_{\rm p}$).} 
\label{fig:ngasonlycombo}
\end{figure*}

\subsection{Key Function} \label{RWI}

In previous studies mentioned in section \ref{intro} of the Rossby Wave Instability, typically the graphics of vorticity or vortensity have been used to indicate the locations of vortices generated by the instability. Additionally, the radial profile of the key function (\citealt{lovelace99}) which incorporates the inverse vortensity and entropy indicates locations of the onset of the RWI through extrema in the profile (a necessary but not sufficient condition for instability). Recently, \cite{chang23} study the stability of dust traps, specifically axisymmetric pressure bumps, to the RWI. They give a new approximate empirical criterion for the RWI based on their results for isothermal, adiabatic and heated gas bumps (see their subsection 3.4). In this study we will focus on presenting the Lovelace key function, as, in addition to presenting a necessary condition for instability, it also provides a useful view on the emergence of vortices, since in our locally isothermal simulations the key function is closely related to the inverse of the vortensity. When the instability triggers, vortices are generated and the extremum in the radial profile of the key function is dampened. This key function is given as, in a gas disc,

\begin{equation} \label{keyfuncgas}
    \mathcal{L}(r) = \frac{\Sigma_{\rm g} S^{2/\gamma}}{2(\nabla \times \textbf{u})_{\hat z}},
\end{equation}

\noindent where $S$ is the entropy given by $S = p_{\rm g}/\Sigma_{\rm g}^{\gamma}$ with $\gamma$ as the adiabatic index. In this study we investigate how effective the key function is in predicting when the instability occurs. However, equation (\ref{keyfuncgas}) has typically been used to present the onset of RWI in gas only simulations of protoplanetary discs and not in dusty discs. In view of the thermodynamic interpretation of \cite{lin17}, it is useful to treat the gas and dust as a single fluid. In order to incorporate the key function into our 2D disc setup of firstly, a locally isothermal equation of state ($\gamma = 1)$ and secondly, with the addition of dust, we simply take the barycentric velocities of the dust and gas phase, given by,

\begin{equation} \label{baryvel}
     \textbf{u}_{\rm b} = \frac{\Sigma_{\rm g}\textbf{u}_{\rm_g} + \Sigma_{\rm d}\textbf{u}_{\rm_d} }{\Sigma_{\rm g} + \Sigma_{\rm d}},
\end{equation}

\noindent with subscripts $\rm g$ and $\rm d$ as the gas and dust respectively and the total density, $\Sigma_{\rm T} = \Sigma_{\rm g} + \Sigma_{\rm d}$, when representing the "mixed" key function given by,

\begin{equation} \label{keyfuncmix}
\begin{split}
    \mathcal{L}^{*}(r) & = \frac{\Sigma_{\rm T}S^2}{2(\nabla \times \textbf{u}_{\rm b})_{\hat z}} \\
    & = \frac{c_s^4(1-f_{\rm d})^2\Sigma_{\rm T}}{2(\nabla \times \textbf{u}_{\rm b})_{\hat z}},
\end{split}
\end{equation} 

\noindent where $f_{\rm d}$ is the dust fraction, $f_{\rm d} = \Sigma_{\rm d}/\Sigma_{\rm T}$ since our vertically integrated pressure can be rewritten as $p_{\rm g} = c_s^2 (1-f_{\rm d})\Sigma_{\rm T}$. We will be using equation (\ref{keyfuncgas}) for plotting the key function for gas only simulations in subsection \ref{kinematics} and equation (\ref{keyfuncmix}) when considering dust and gas disc simulations in the remaining subsections of our results. 

In the thermodynamic view of \cite{lin17}, equation (\ref{keyfuncmix}) gives a necessary condition for the RWI to develop in a perfectly coupled mixture of dust with an isothermal gas. A varying dust fraction introduces entropy gradients in the single fluid formalism, and hence the stability properties of the mixture can differ from that of the gas only. In the single fluid model (\citealt{laibe14}; \citealt{lin17}) the interaction between the gas and dust is taken into account through a cooling term due to the drag forces. If this approach was used in the study of the RWI, it would correspond to a gas RWI with cooling. In this approach care must be taken to disentangle the RWI from drag instabilities as both play a important role in unstable disc evolutions.

\section{Results} \label{results}

We present our results in this section comparing the evolution of a gas only disc and ring generated by a Neptune sized planet with one that includes dust. In section \ref{kinematics} we review the effect of the kinematic viscosity on the pressure bump and ring generated by the planet for gas only simulations. Section \ref{lowvisc} shows the impact of dust to total ratio of $1:100$ in a low viscosity disc with well coupled grains with $\stokes=0.01$. We study how the dust fraction and Stokes number can effect the evolution of the ring in section \ref{stability}.

In order to show that the instability of the ring is not caused by effects related to gas drag we include in section \ref{cutoffsec} results where we force perfect coupling between gas and dust at the point where the ring would go unstable, and since the instability proceeds as normal, we conclude that drag does not play a significant role in making the dust ring unstable. Finally in subsection \ref{longvort} we investigate the vortices generated by dusty RWI and briefly present the possible long term evolution and show the possible implications.

\subsection{Stability of Gas Rings} \label{kinematics}

For our test cases we ran benchmark simulations of gas only discs with an embedded Neptune sized planet, while varying the uniform kinematic viscosity, $\nu \in [10^{-5}, 10^{-6}, 10^{-7}]$. We use a resolution of (768,1536) cells in the radial and azimuthal direction respectively and the simulation time for each case is 100 orbits. We ran these simulations to review how the kinematic viscosity can inhibit the requirements for the RWI to trigger and find a a setup which would be useful to study how the addition of dust affects the onset of the RWI. In the inviscid limit, \cite{devalborro06} showed the appearance of vortices through the RWI for this setup.

In Fig. \ref{fig:ngasonlycombo} we plot the density of the gas for the varying kinematic viscosities. After 100 orbits we can see that for the most viscous case (left panel), the Neptune sized planet has carved a shallow gap in the disc and faint rings are generated by the planet outside the gap in the pressure bumps. The expected spiral density waves are launched from the planet's location through the disc (\citealt{goldreich79}; \citealt{goldreich80}). For longer time integration, the morphology of the disc remains stable as the viscous damping is large enough to suppress conditions for instabilities. Moving onto the middle panel of Fig. \ref{fig:ngasonlycombo}, for the $\nu = 10^{-6}$, we see a deeper gap carved as the kinematic viscosity of the disc is lessened and the gas ring is more pronounced at the gap edges ($r = 0.9, 1.1 r_{\rm p}$) than the first. In this case we note that the gas ring shows instabilities by 200 orbits as gradually the density gradient of the gap edge carved out by the planet is extreme enough to trigger the RWI, however these are quickly suppressed by the viscosity. In the right panel of Fig. \ref{fig:ngasonlycombo} with the lowest kinematic viscosity we see that the gap edge created has triggered the RWI within 100 orbits and vortices have been generated interior and exterior to the planet. These vortices can be identified as gas overdensities, and alter the region of the pressure bump and ring generated by the planet through their own interactions of merging together in the azimuthal direction. One important question introduced then, which will be explored briefly in the last subsection of the results, is "What is the long term evolution of the vortices exterior to a gap carving planet?". As isolated vortices are subject to viscous damping (\citealt{fu14}) while a planet carving a deep enough gap in the disc can trigger the RWI into generating vortices, this would be key to the evolution and possible further planetesimal and planetary formation. Furthermore, the addition of dust into a vortex can affect their lifetimes and evolution due to instabilities in the vortices, introduced or enhanced by the presence of dust (e.g. \citealt{lovascio22}).

\begin{figure}
\includegraphics[width=\linewidth]{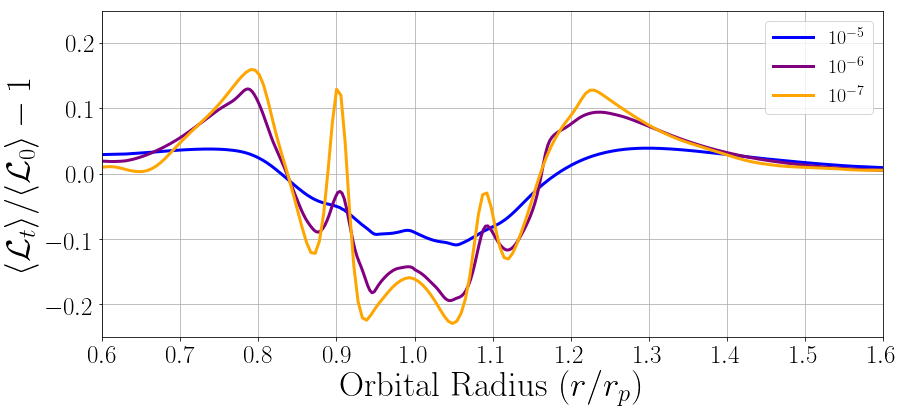}
\caption{Radial profile of the azimuthally averaged key functions for $\nu \in [10^{-5}, 10^{-6}, 10^{-7}]$. The lower viscosity setups have a larger peak at the location of the gap edges ($r = 0.9 r_{\rm p}, 1.1 r_{\rm p}$) which lead to development of the RWI. } 
\label{fig:keycombo}
\end{figure}

\begin{figure}
\includegraphics[width=\linewidth]{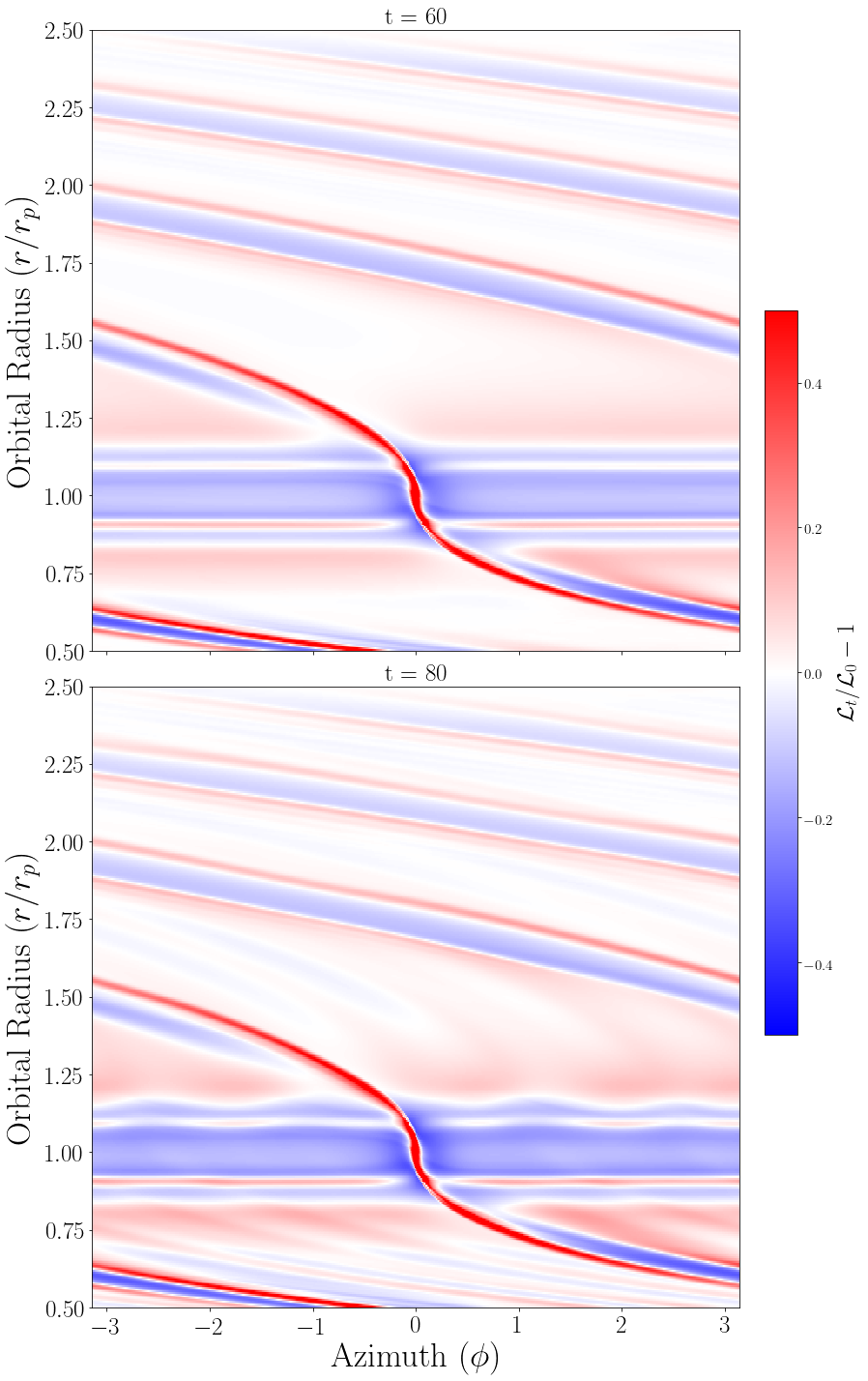}
\caption{Key function plot for a gaseous disc with Neptune sized planet embedded at $1 r_{\rm p}$ after $t = 60, 80$ orbits. The kinematic viscosity is, $\nu = 10^{-7}$ and the gap edges interior ($r = 0.9 r_{\rm p}$) and exterior ($r = 1.1 r_{\rm p}$) to the planet has triggered the RWI generating a number of vortices which are more pronounced exterior to the planet.} 
\label{fig:timecombo}
\end{figure}

\begin{figure}
\begin{subfigure}{\linewidth}
    \centering
    \includegraphics[width=\linewidth]{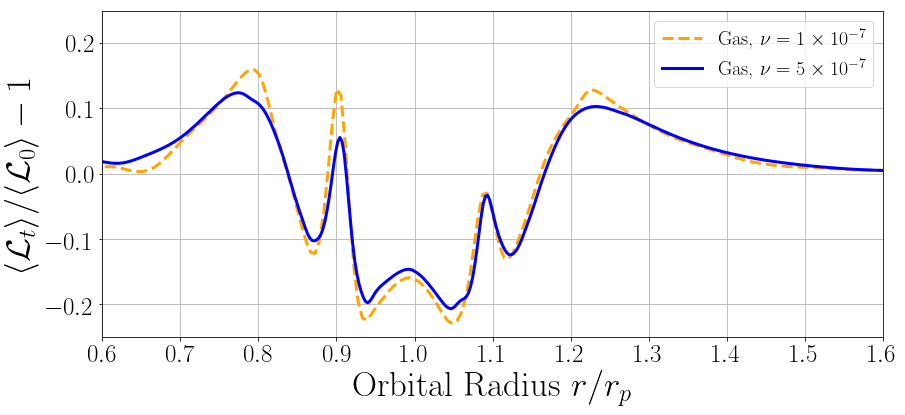}
\end{subfigure}
\begin{subfigure}{\linewidth}
    \centering
    \includegraphics[width=\linewidth]{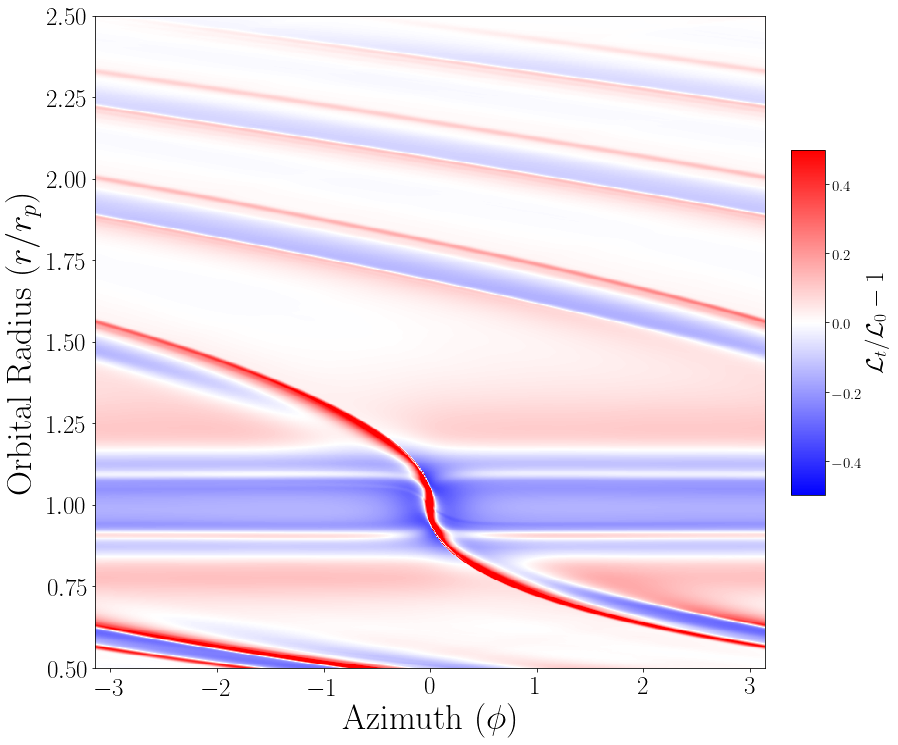}
\end{subfigure}
\caption{Radial profile of the azimuthally averaged key function (top) for stable (solid blue line) and unstable (dotted orange) gaseous discs. 2D key function plot (bottom) for a gaseous disc with Neptune sized planet embedded at $1 r_{\rm p}$ after $t = 100$ orbits. The kinematic viscosity is set to $\nu = 5\times10^{-7}$. In this case the kinematic viscosity is large enough to suppress the onset of the RWI during its evolution as shown in the bottom plot with a smooth gap edge interior and exterior to the planet, however it is not as indicative in the radial profile of the key function compared to the lower viscosity case which goes unstable.} 
\label{fig:gas5e-7}
\end{figure}

Fig. \ref{fig:keycombo} presents the radial profile of the azimuthally averaged key function at $t = 100$ orbits over the initial profile of the key function in the disc. Initially with a kinematic viscosity of $10^{-5}$, we see that in the density plot the evolution of the disc and gap is very stable with shallow rings interior and exterior to the shallow gap. This is indicated in the key function as well since the rings do not go unstable and there are no extremum in the radial profile of the key function which is highlighted by the blue line. We see that as the kinematic viscosity is lowered to $10^{-6}$, after 100 orbits peaks are developed in the key function at the location of the gap edges ($r = 0.9, 1.1 r_{\rm p}$) and are much more pronounced for the case of the lowest viscosity of $10^{-7}$. Since the RWI has already developed and generated vortices after 100 orbits, the peaks are already at a lower point than when it first triggers the instability. However it is clear that a sharp bump or extremum in the key function indicates locations where the RWI can occur especially when comparing the highest and lowest viscosity cases. 

We have included 2D plots of the key function in Fig. \ref{fig:timecombo} for the low viscosity run ($\nu=10^{-7}$). The reason for this is that, in particular in the gas-only case, the key function is essentially the inverse of the vortensity and therefore useful in showing the development of the RWI and the associated vortices. Note that while in the gas-only case, the entropy factor in the key function only plays a minor role (due to the locally isothermal nature of the setup), the inclusion of dust affects the entropy part of the key function due to its dependence on the dust fraction in equation (\ref{keyfuncmix}). The key function plots presented show the evolution of the disc from 60 to 80 orbits for $\nu=10^{-7}$. We see that the gap edges interior and exterior have triggered the RWI, generating at least four vortices exterior to the outer gap edge. This in turn not only destabilises the gas ring in the pressure bump but causes additional weaker spiral arms to propagate throughout the disc from the vortices' locations which is clearer in Fig. \ref{fig:timecombo}. Continuing on from 80 orbits these smaller vortices would merge together to form bigger vortices as seen in Fig. \ref{fig:ngasonlycombo} where only two can be seen. This process reduces the amplitude of the peak in terms of its radial profile of the key function but the extremum is still visible after 200 orbits.

Based on the test cases so far, we chose to investigate two different cases, based on whether the RWI develops within 100 orbits in the gas-only setup. First of all we will consider a setup of the disc where the RWI may appear within 100 orbits with a low kinematic viscosity of $10^{-7}$. After that, we consider discs of higher viscosity and study whether the addition of dust makes the disc more prone to instability. Note that in particular in the intermediate case of $\nu=10^{-6}$, it is important to establish whether the addition of dust speeds up or delays the onset of the RWI.

In order to establish a more precise limit on the viscosity for which the gas disc remains stable for 100 orbits, we also present the case of $\nu = 5\times 10^{-7}$. This case will be used in the following sections, and falls in between the cases $\nu=10^{-7}$ and $\nu=10^{-6}$ of Fig. \ref{fig:ngasonlycombo}. The key function for this simulation is shown in Fig. \ref{fig:gas5e-7}, both azimuthally averaged (top panel) and the full 2D structure (bottom panel). From the bottom panel, it is clear that after 100 orbits no vortices have formed, and while the RWI may be triggered at later times, after 100 orbits the gas disc looks quiet and stable. The peaks in the 1D key function are comparable to those in the (unstable) $\nu=10^{-7}$ case, but note that the latter will have been affected by the emerged vortices.

\subsection{Dust in an RWI Unstable Gas} \label{lowvisc}

\begin{figure}
\begin{subfigure}{\linewidth}
    \centering
    \includegraphics[width=\linewidth]{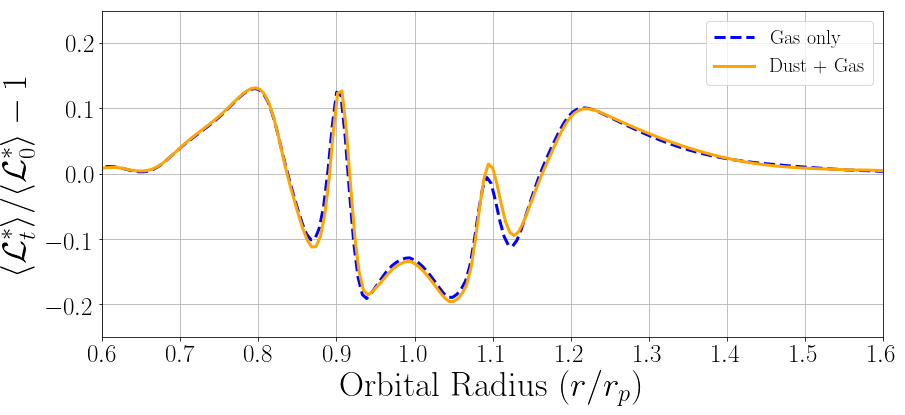}
\end{subfigure}
\begin{subfigure}{\linewidth}
    \centering
    \includegraphics[width=\linewidth]{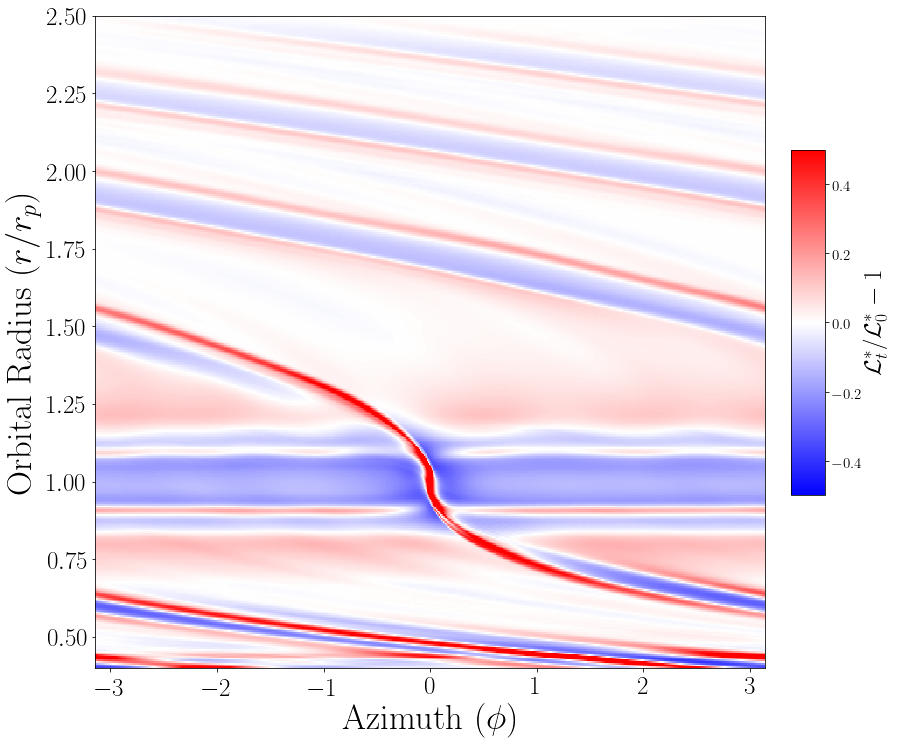}
\end{subfigure}
\caption{Radial profile of the azimuthally averaged "mixed" key function (top) for dust and gas (solid orange) and radial profile of gas only key function (dotted blue). 2D key function plot (bottom) for the dust and gas disc with Neptune sized planet embedded at $1 r_{\rm p}$ after $t = 80$ orbits. The dust fraction is set to $f_{\rm d} = 0.01$, Stokes number, $\stokes = 0.01$ and kinematic viscosity, $\nu = 10^{-7}$. With the addition of dust the gap edges interior and exterior still trigger the RWI which generates vortices but has little differences compared to the gas only case. } 
\label{fig:dustylinecombo}
\end{figure}

We first set the kinematic viscosity to $\nu=10^{-7}$, a case where the gas-only RWI develops within 100 orbits, and investigate whether adding dust makes a noticeable change in the evolution of the disc compared to the gas only case. We test how the evolution of the disc may change with a typical dust to total ratio of 1:100. The Stokes number we use for the first test is $\stokes=0.01$ which typically corresponds to smaller grain sizes or strong drag of the gas onto the dust, making it tightly coupled. We present the plots of the azimuthally averaged and 2D key function of this simulation after 80 orbits in Fig. \ref{fig:dustylinecombo}. 

\begin{figure}
\includegraphics[scale=0.26]{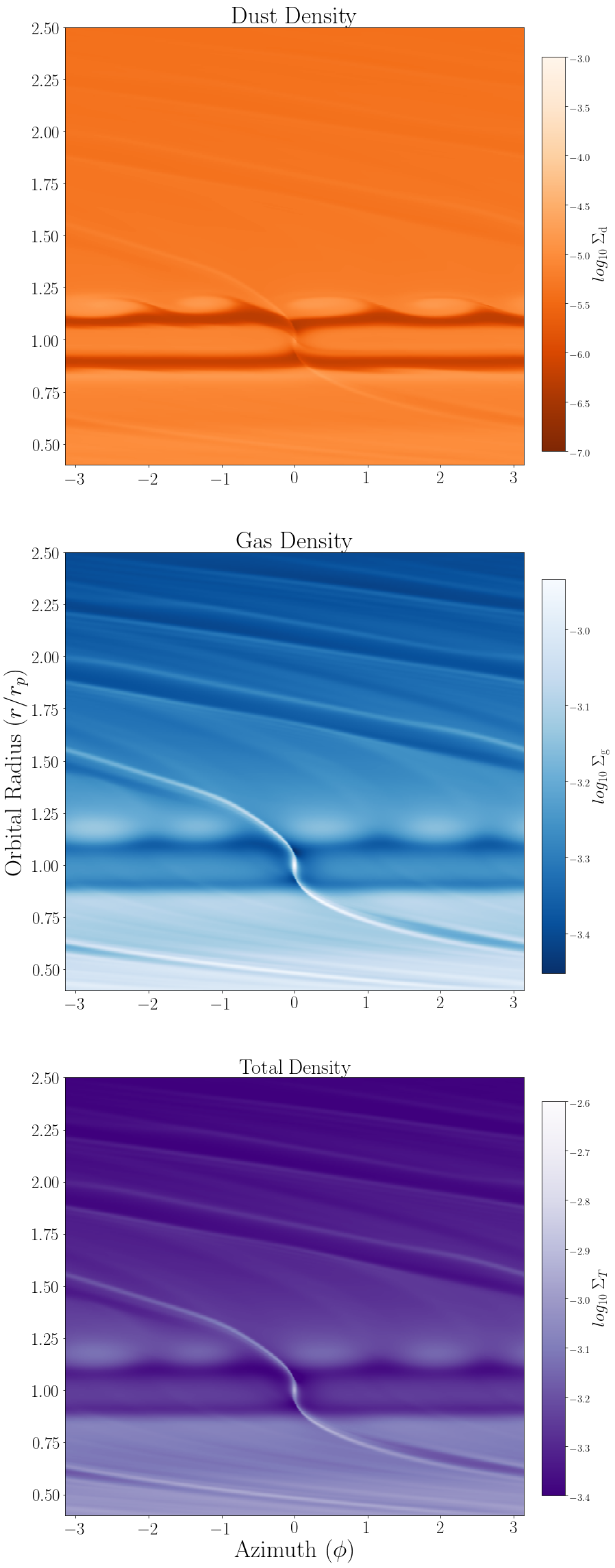}
\caption{2D density plots for the dust (top), gas (middle), and total (bottom) of a Neptune planet embedded disc at $t = 90$ orbits. The dust fraction is set to $f_{\rm d} = 0.01$, Stokes number, $\stokes = 0.01$ and kinematic viscosity, $\nu = 10^{-7}$. We see that the dust from the outer ring is accumulated quickly into the vortices generated by the gap edge triggering the RWI.} 
\label{fig:001st001fd_dens}
\end{figure}

\begin{figure}
\includegraphics[scale=0.26]{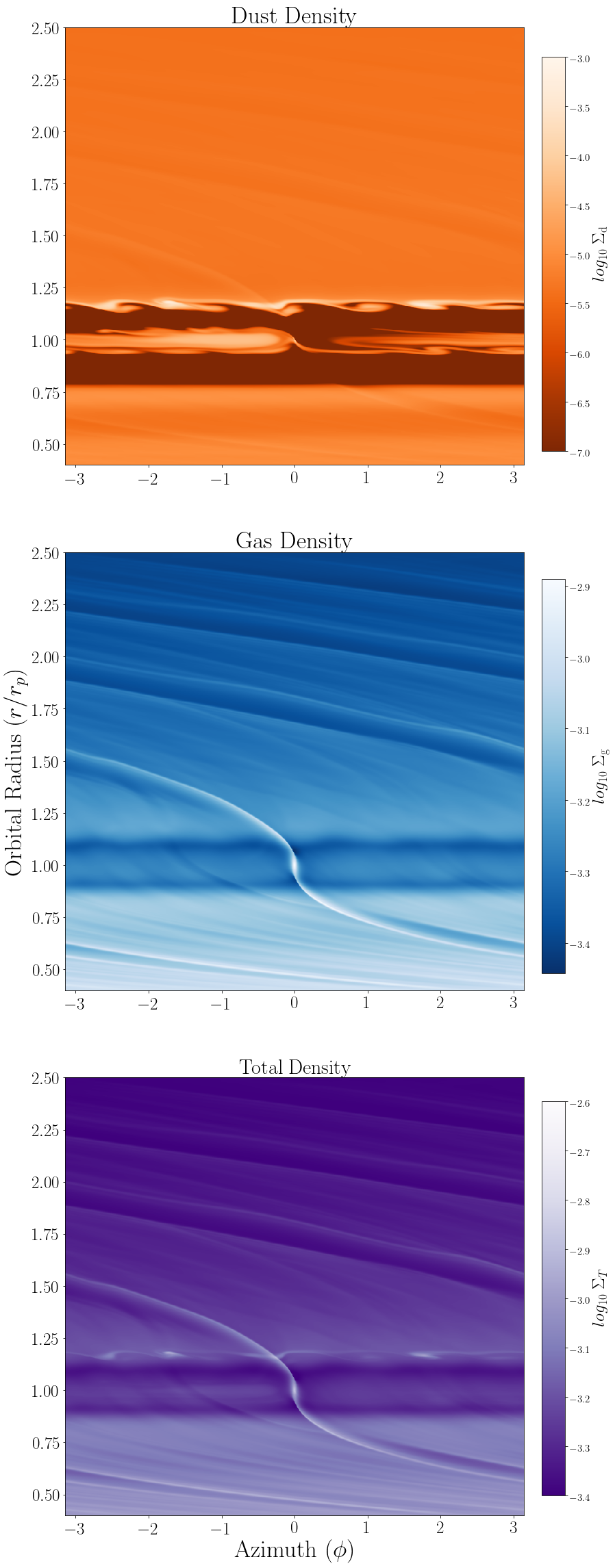}
\caption{2D density plots for the dust (top), gas (middle), and total (bottom) of a Neptune planet embedded disc at $t = 90$ orbits. The dust fraction is set to $f_{\rm d} = 0.01$, Stokes number, $\stokes = 0.1$ and kinematic viscosity, $\nu = 10^{-7}$. The thin dusty ring built exterior is destabilised by vortices at the gap edge and material is accumulated into vortex centres which causes weak spiral density waves to be launched from the vortices.} 
\label{fig:01st001fd_dens}
\end{figure}

From the plots we see that after 80 orbits the evolution of the disc with dust added are very similar to the gas only case when plotting the "mixed" key function. The radial profile of the azimuthally averaged key function has the same features as in Fig. \ref{fig:keycombo} with a larger peak at the inner gap edge and smaller peak at the outer gap edge. The exact values at the peaks and troughs for the gas only and "mixed" key function can not be directly compared, however the features themselves in the "mixed" key function plots show good indication of places prone to the RWI. In terms of the 2D plot we see the similar features of the gap edges triggering the RWI and generating vortices outside the gap edges with minor differences. The vortices generated by the unstable gas quickly destabilise the dusty ring building up exterior to the planet, accumulating the material into the vortices. In Fig. \ref{fig:001st001fd_dens} we show the density plots of the dust, gas and total, ten orbits later at $t = 90$. We see that the outer ring is broken up by the four vortices while the dusty ring interior to the planet remains stable but undergoes the same evolution by 100 orbits when the inner gap edge triggers the RWI. In this case we see that the gas RWI is largely unaffected by the presence of dust, when comparing the key function plots, as it is swept into the vortices. We note however that the long term evolution of the disc is expected to be different as the vortices are loaded with more dust.

When we increase the Stokes number to 0.1, a similar evolution is observed however in this case, the dusty ring formed exterior to the planet is much thinner in the radial direction before the vortices destabilise the ring and accumulate the material into the vortex centres. We present the density plots in Fig. \ref{fig:01st001fd_dens} where we see that the material built up in the vortex cores after 90 orbits start to launch spiral density waves throughout the disc as dust builds up. When observing the gas phase for this setup, we can see that by 90 orbits the vortices are not as defined as in the previous case of $\stokes = 0.01$ as more dust is loaded into the vortex cores. A reason for this would be drag instabilities occuring in the vortices between the dust and the gas. Therefore in this case, we do see a larger difference between the evolution of the vortices generated by the triggering of the RWI at the gap edge for a gas only and dust+gas disc. However we do not observe if the dust makes the disc less prone to the RWI within 100 orbits as the RWI is triggered by the gas at an early stage in both scenarios of adding dust to a disc. In these cases, the gas RWI dominates the early evolution of the disc and the dust does not contribute significantly to the triggering of the instability.

As $10^{-7}$ for the kinematic viscosity is typically a very low viscosity value compared to observations and current upper constraints in some discs (\citealt{rosotti23}), it is important to note that regions of low viscosity can occur in discs such as the dead zones (\citealt{gammie96}; \citealt{armitage11}).

\subsection{Dust in an RWI Stable Gas} \label{stability}

Following on, we investigate if there is a point where the dust can make a noticeable difference in the evolution of the disc for a stable gas and whether it makes the gap edges more prone to the RWI or if the dust buildup could source a new location for the RWI to occur. From our test case of $f_{\rm d} = 0.01$, $\stokes = 0.01$ and $\nu = 10^{-7}$ we varied the three parameters to find if there was a defining factor to triggering the onset of the RWI. As seen before, the kinematic viscosity plays an important part in suppressing the instability therefore for the next results we use a kinematic viscosity of $\nu = 5\times10^{-7}$ to compare with our gas only results in Fig. \ref{fig:gas5e-7} where the evolution of the region exterior to the planet's outer gap edge remained stable.

\subsubsection{Varying the Dust Fraction} \label{dustfracs}

\begin{figure}
\includegraphics[width=\linewidth]{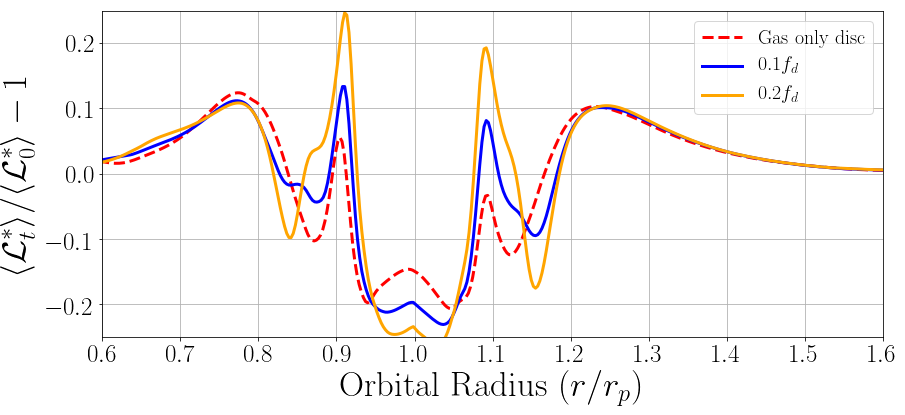}
\caption{Radial profile of the azimuthally averaged "mixed" key functions for a dust and gas disc with dust fractions of $f_{\rm d} = 0.1, 0.2$ and Stokes number, $\stokes = 0.01$. The kinematic viscosity is set to $\nu = 5\times10^{-7}$ and the evolution of the disc has reached $t = 100$ orbits. We see that large peaks have formed at the location of the gap edges showing the growth of the "mixed" key function compared to initial.} 
\label{fig:dustfractions}
\end{figure}

With a focus on $\stokes=0.01$, we found that our standard dust fraction $f_{\rm d}=0.01$ made no difference to the stability of the disc within 100 orbits. We now investigate whether larger dust fractions, where the feedback is more important, changes this situation. In Fig. \ref{fig:dustfractions} we present the azimuthally averaged "mixed" key function for discs with dust fractions, $f_{\rm d} = 0.1, 0.2$. 

\begin{figure}
\includegraphics[width=\linewidth]{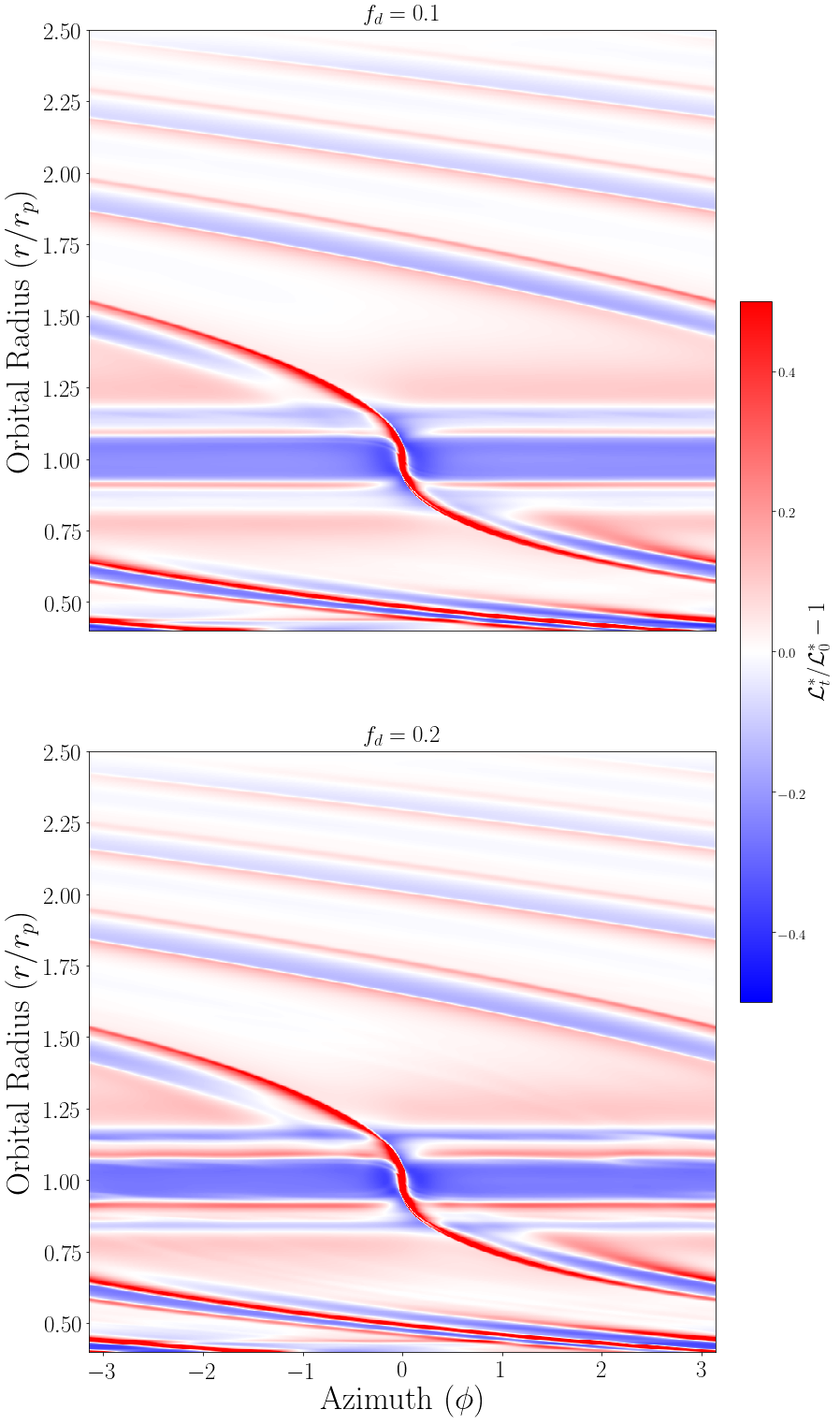}
\caption{"Mixed" key function plot for dust and gas disc with dust fractions, $f_{\rm d} = 0.1$ (top) and $0.2$ (bottom) with Stokes number, $\stokes = 0.01$ for both cases. The evolution of the disc has reached $t = 100$ orbits. The gap edges interior and exterior to the planet remain generally stable during its evolution with the dusty rings being largely undisrupted.} 
\label{fig:timecombo2}
\end{figure}

We see that after 100 orbits, the peaks formed at the gap edges are much larger than previous radial profiles of the "mixed" key function especially comparing to Fig. \ref{fig:dustylinecombo} which had a smaller dust fraction of $0.01$ and lower kinematic viscosity. Normally with extremum in the radial profile, we would expect the RWI to trigger but the gap edges still remained stable. This is shown in Fig. \ref{fig:timecombo2} where we present the 2D plot of the "mixed" key function. In both cases, the gap edges remain stable and smooth generally. In the second case of $f_{\rm d} = 0.2$ we see that there is a hint of an instability occurring soon or being suppressed at the gap edges near the azimuthal coordinate $\phi \approx 2.5$, however it is not clear enough to discern during the evolution of the disc. In terms of the density, the dust build-up in the ring interior and exterior to the planet reached a ratio of around $1:2$ of dust to gas and the total density in the exterior ring was larger than in the gas only case, causing a steeper gradient between the gap edge and ring which should indicate a higher susceptibility to the RWI triggering. The initial dust fraction may play a part in how prone the gap edges are to the RWI, as dust may build-up in the rings that would generate a steep enough gradient and more should be explored in further studies however we can see in this setup, that increasing the initial dust fraction of the disc up to $f_{\rm d} = 0.2$, that the gap edges are not more prone to the RWI after $100$ orbits.

We note that the azimuthally averaged "mixed" key function so far may indicate possible locations of the RWI triggering but the amplitude of the peaks compared to initial values does not necessarily mean that the RWI has triggered already or is more prone to it during early evolution. It could be possible that with the inclusion of dust, even though peaks in the radial profile of the "mixed" key function are larger, the RWI may be suppressed. Perhaps for even larger dust fractions, the RWI can be triggered, but such a large dust fraction is unlikely to occur on a global scale. It is probably more realistic to let a larger dust fraction occur locally in a pressure bump, which can be achieved on reasonable time scales for larger Stokes numbers. This is what we do next.

\subsubsection{Stokes Number} \label{stokes}

\begin{figure}
    \centering
    \includegraphics[scale=0.26]{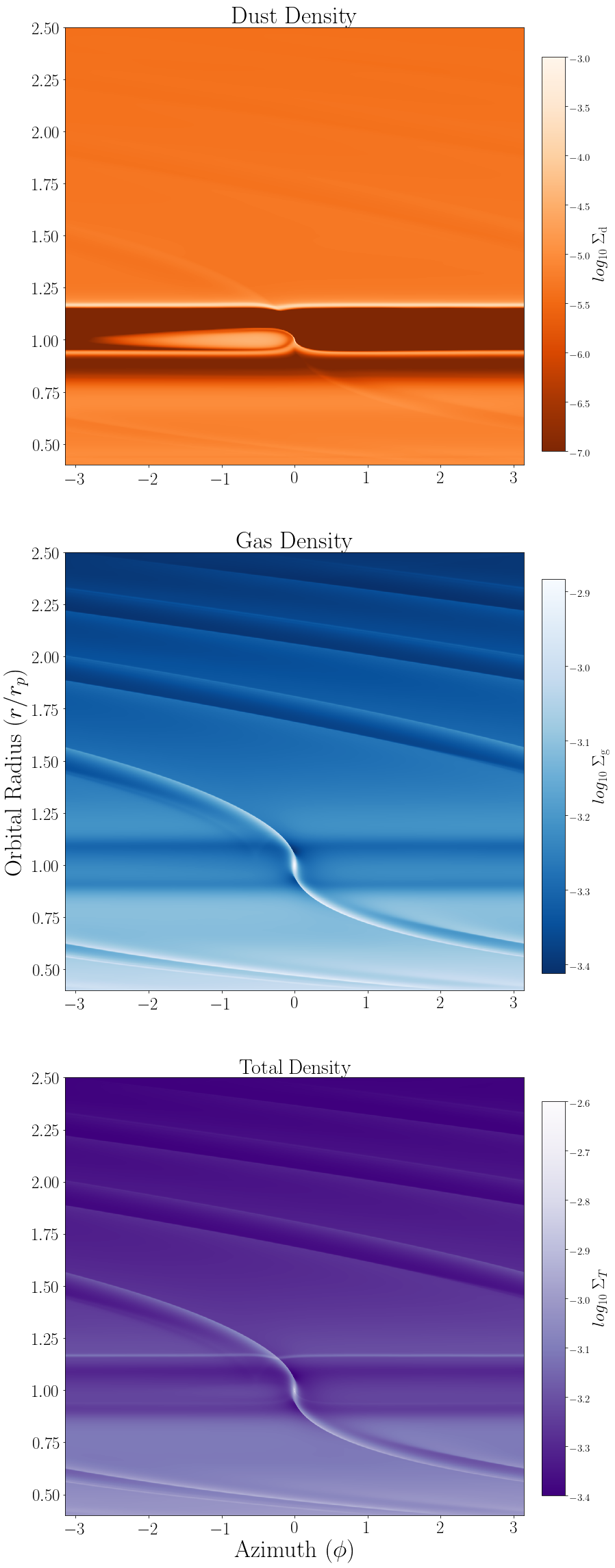}
\caption{Density plots for dust (top left), gas (top right) and total (bottom) of the Neptune planet embedded disc with dust fraction, $f_{\rm d} = 0.01$ and Stokes number, $\stokes = 0.2$. The evolution of the disc is at $t = 60$ orbits. We see a smooth thin dust ring has accumulated exterior to the planet and the features are currently stable.}
\label{fig:t60dens}
\end{figure}

\begin{figure}
    \centering
    \includegraphics[scale=0.26]{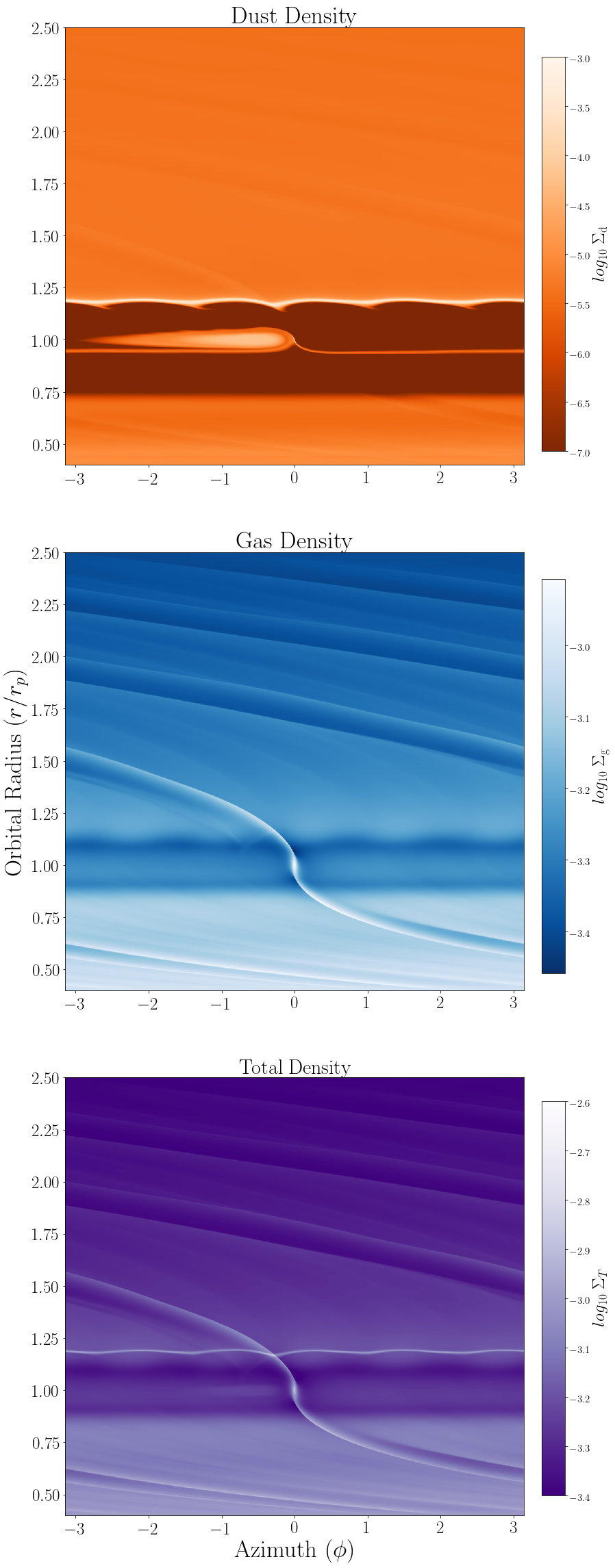}
\caption{Density plots for dust (top left), gas (top right) and total (bottom) of the Neptune planet embedded disc with dust fraction, $f_{\rm d} = 0.01$ and Stokes number, $\stokes = 0.2$. The evolution of the disc is at $t = 100$ orbits. The dust ring has triggered the RWI at the rings location and generated vortices which also affect the gas phase and not at the gap edge as previous cases.}
\label{fig:t100dens}
\end{figure}

Having shown that increasing the dust fraction up to $0.2$ had little effect on how prone the regions near the planet were to the RWI within $100$ orbits, we increased the Stokes number of the dust but kept the original dust fraction of $0.01$. For this, we first highlight the simulations with a Stokes number, $\stokes = 0.2$, with the illustrations of Fig. \ref{fig:t60dens} and Fig. \ref{fig:t100dens} which show the dust and gas density plots and the combined total density at different timesteps. In Fig. \ref{fig:t60dens}, we show the density plots of the disc at $t = 60$ orbits. When comparing just the gas density plot in this figure and the morphology of the key function of the gas only control case in Fig. \ref{fig:ngasonlycombo}, we see that both are similar in terms of stability of the gap edges and the expected shallow gap and spiral density waves launching from the planet's location. In terms of the dust we see that a thin outer ring has built up in the pressure bump ($r = 1.2 r_{\rm p})$ exterior to the planet past the gap edge. In the radial profile of the dust density we observe this as a sharp peak compared to section \ref{dustfracs} of increasing the dust fraction where the dust accumulated into a wide bump. This, we found to be important as we present the next figure which shows the evolution of the disc at $100$ orbits.

\begin{figure}
    \centering
    \includegraphics[width=\linewidth]{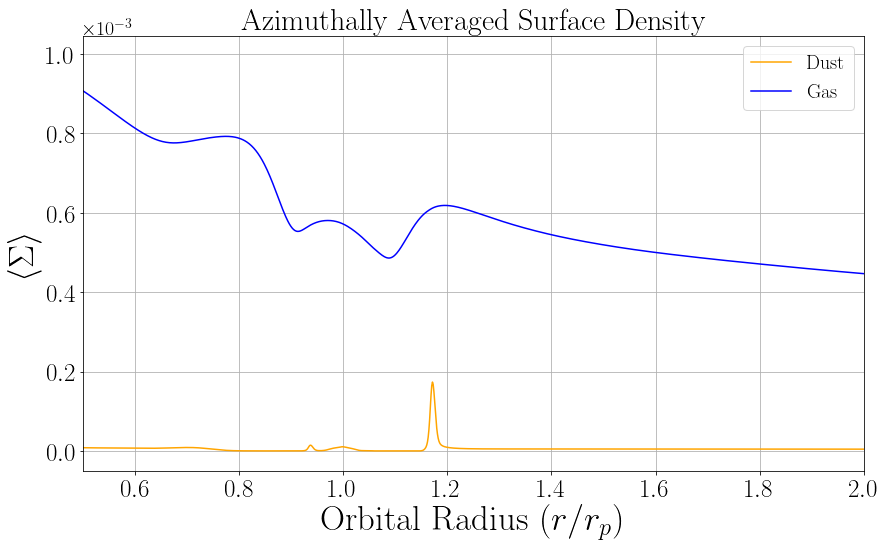}
\caption{Radial profile of the azimuthally averaged surface densities for the dust (orange) and gas (blue) after $t =70$ orbits. The dust evolution shows a sharp peak building at the outer pressure bump before the instability is triggered at around $t = 78$.}
\label{fig:dustgasazi}
\end{figure}

\begin{figure}
\begin{subfigure}{\linewidth}
    \centering
    \includegraphics[width=\linewidth]{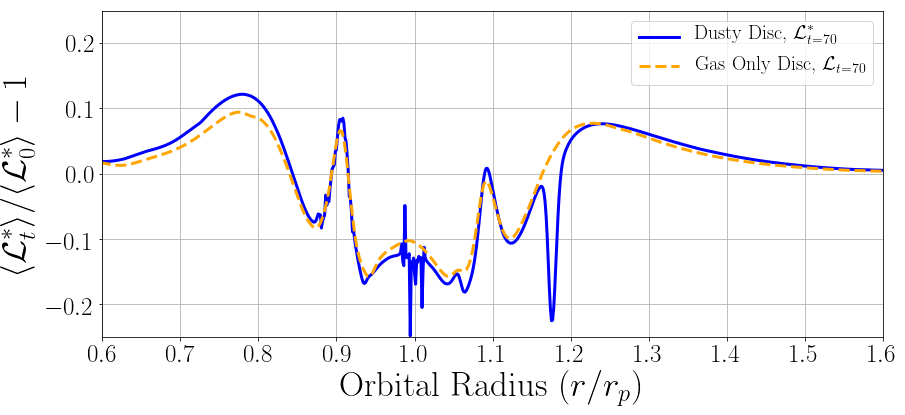}
\end{subfigure}
\begin{subfigure}{\linewidth}
    \centering
    \includegraphics[width=\linewidth]{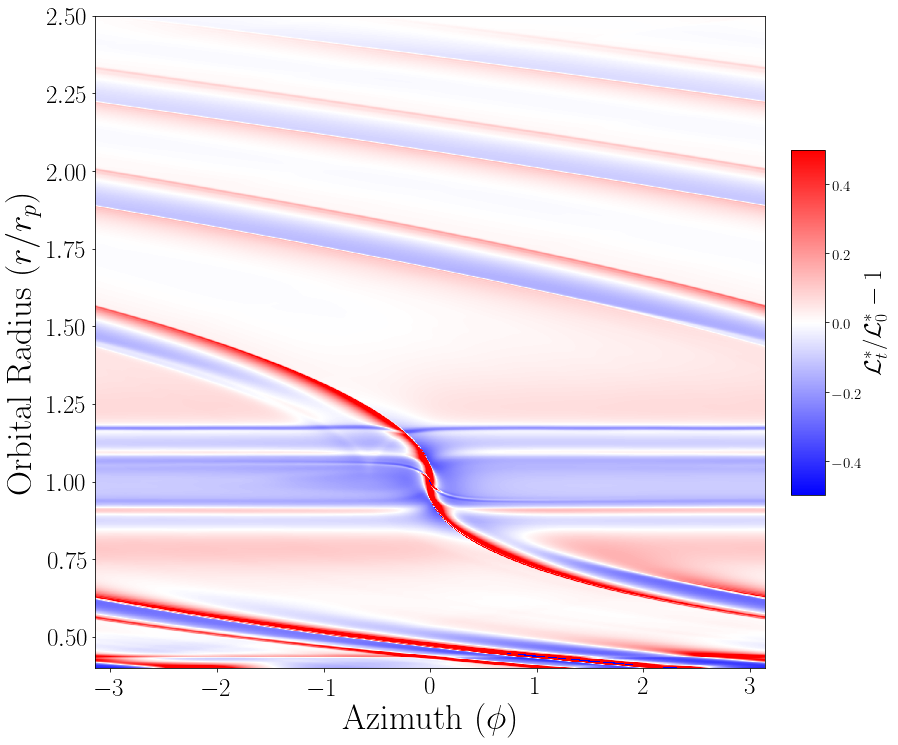}
\end{subfigure}
\caption{Radial profile of the azimuthally averaged "mixed" key function of a dust and gas disc (blue line) compared to gas key function for a gas only disc (dashed orange line) (top plot) and 2D key function plot (bottom) for a dust and gas disc with Neptune sized planet embedded at $1 r_{\rm p}$ after $t = 70$ orbits. The dust fraction is set to $f_{\rm d} = 0.01$, Stokes number, $\stokes = 0.2$ and kinematic viscosity, $\nu = 5\times 10^{-7}$. The radial profile shows a sharp minimum at the dust ring's location and indicates that the width of the extremum is an important factor which indicates location of regions prone to the RWI.}
\label{fig:t70key}
\end{figure}

In Fig. \ref{fig:t100dens}, we see that the RWI has triggered and vortices have been generated at the location of the outer ring. The thin ring of dust that had built up in the pressure bump is distorted by the vortices and the dust itself is being accumulated in the centres of the vortices creating a wave shaped ring. Looking at the gas density plot, we see that the gas phase itself has become unstable exterior to the planet. Comparing this with our control case of a gas only disc we show that the addition of dust can have a significant effect on how prone the disc can be to the RWI. Surprisingly, the cause of the RWI in this case is not due to the gradient near the gap edges but due to the sharp peak in the outer ring. This is to say that the dust phase has triggered the RWI due to the contribution of the sharp dust density peak accumulating in the middle of the gas ring that forms in the outer pressure bump and causes the region to become unstable, generating vortices which are visible in both the dust and the gas even though the disc only comprises of $1\%$ dust.

\begin{figure}
    \centering
    \includegraphics[width=\linewidth]{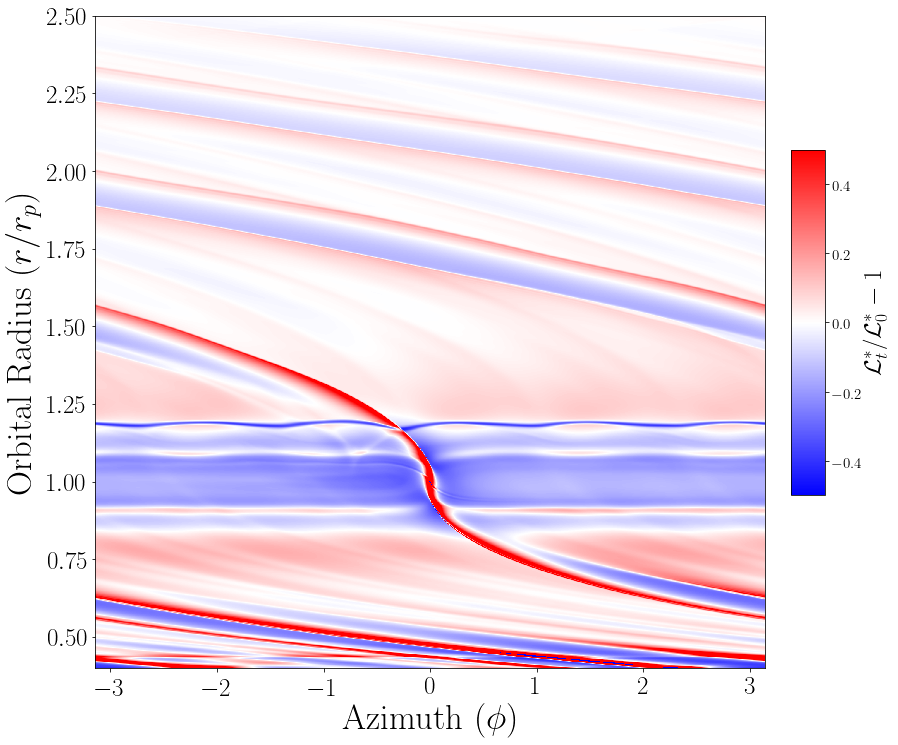}
\caption{2D "mixed" key function plot at $t = 100$ orbits for a dust and gas disc with Neptune sized planet embedded disc. The dust fraction is set to $f_{\rm d} = 0.01$, Stokes number, $St = 0.2$ and kinematic viscosity $\nu = 5\times10^{-7}$.} 
\label{fig:t100key}
\end{figure}

From our data, we find that the RWI triggers at around $t = 78$ orbits. Therefore we analyse and present the next figures from the data at $t =70$ which is before the instability triggers. Firsly we present the azimuthally averaged surface densities for the dust and gas in Fig. \ref{fig:dustgasazi}. We note that the sharp dust peak in the density profile of the disc at $t =70$ reaches a value that is below 1:3 for the dust to gas ratio at the outer pressure bump compared to the wide dust bump build-up for the higher dust fraction case, $f_{\rm d} = 0.2$, which reached a dust to gas ratio of around 1:2. Additionally we point out that the maximum azimuthally averaged total density value for the peak of the outer ring in the increased dust fraction case is larger than this case of a higher Stokes number. This indicates that the amplitude of the extremum in the total density profile is less indicative of the RWI triggering compared to the width of the extremum which is shown as a sharp peak in the higher Stokes number case. 

In Fig. \ref{fig:t70key} we present the "mixed" key function plots for the higher Stokes number case at $t =70$. Firstly we point out that there are sharp lines at the planets location and the inner minimum at around $r = 0.87r_{\rm p}$ that are temporary (lasting one or two timesteps) and we believe are due to the dust and gas phase interacting more erratically in shock regions compared to the more tightly coupled dust in the previous case of higher dust fraction. This can be seen more in the 2D plot on the edges of the spiral density wave launched by the planet. Higher resolution simulations would most likely remove the temporary spikes therefore we disregard them when evaluating the key function plots. 

The azimuthally averaged "mixed" key function plot (top of Fig. \ref{fig:t70key}) shows the expected peaks at the gap edges which have been seen in all previous plots of the radial profile of the key function located at $r = 0.9$ and $1.1 r_{\rm p}$. As shown previously in this section, the gap edges themselves do not trigger the RWI in the gas only case and higher dust fraction cases. In addition to the peaks found at the gap edges we also see an extremum at around $r = 1.17 r_{\rm p}$ which is instead a minimum in the radial profile. We can see the minimum does not exist in the radial profile of the key function for the gas only disc. Since the addition of dust causes an entropy decrease, the build up of dust at the outer bump creates the minimum in the key function. We see that this minimum is sharper than the local maxima at the gap edges which reinforces our previous statement of the width of the extremum in the density profile and key function is more indicative of conditions for the RWI triggering rather than just the ampltiude as it would provide a steep gradient for the RWI to trigger. The 2D "mixed" key function plot (bottom of Fig. \ref{fig:t70key}) shows the location of the narrow ring-like structure which is not present for the gas only case in Fig. \ref{fig:gas5e-7} and the higher dust fraction plots in Fig. \ref{fig:timecombo2}. Additionally, we present the evolution of the "mixed" key function at $t = 100$ in Fig. \ref{fig:t100key}. From the figure, we see that the vortices generated by the RWI start to destabilise the dusty ring as we expect the material to be accumulated into the vortex centres. A further analysis of the vortices and comparison with gas only RWI will be discussed in subsubsection \ref{initcomp}.

Overall we have shown that the addition of dust in a gaseous disc can alter the evolution of the outer ring significantly, but only for relatively large Stokes numbers. In this case, the RWI is triggered by the thin dusty ring formed which generates vortices that disrupt the observable dusty ring. This has important implications for the future evolution of the dusty ring and the disc morphology as we expect the vortices to accumulate the dust from the ring location, effectively dismantling the substructure and leading to possible planetesimal or planetary formation. We explore this implication further in subsection \ref{long} for the long term evolution.

\subsubsection{Growth of the Instability}

\begin{figure}
    \centering
    \includegraphics[width=\linewidth]{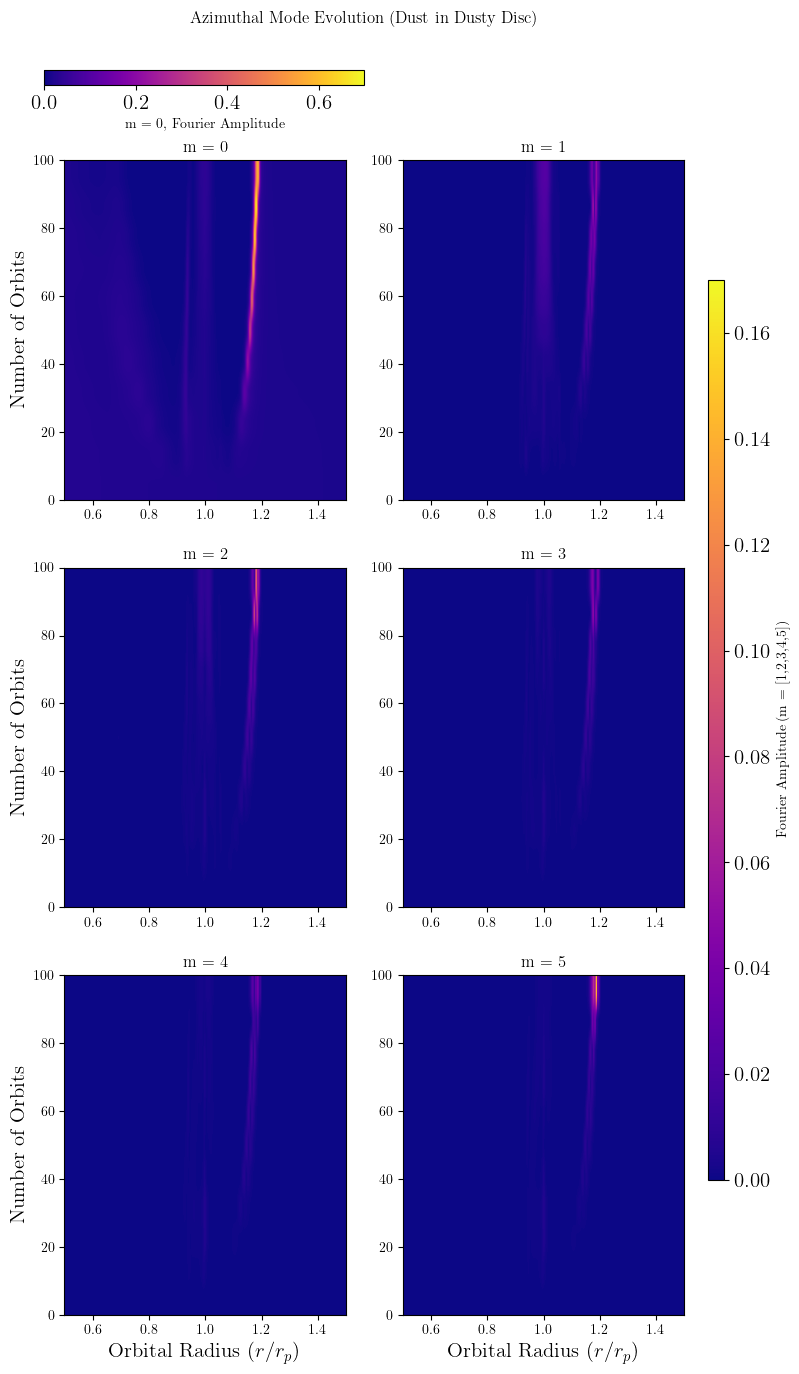}
\caption{Time evolution of the Fourier amplitudes of modes, $m = 0$ to $5$, for the dust in a Neptune planet embedded dusty disc with dust fraction $f_d = 0.01$, Stokes number, $St = 0.2$ and $\nu = 5 \times 10^{-7}$. The growth of the Fourier amplitudes at $r/r_p = 1.2$ dominates where the location of the thin dusty ring is formed and triggers the RWI.} 
\label{fig:dust_growth}
\end{figure}

\begin{figure}
    \centering
    \includegraphics[width=\linewidth]{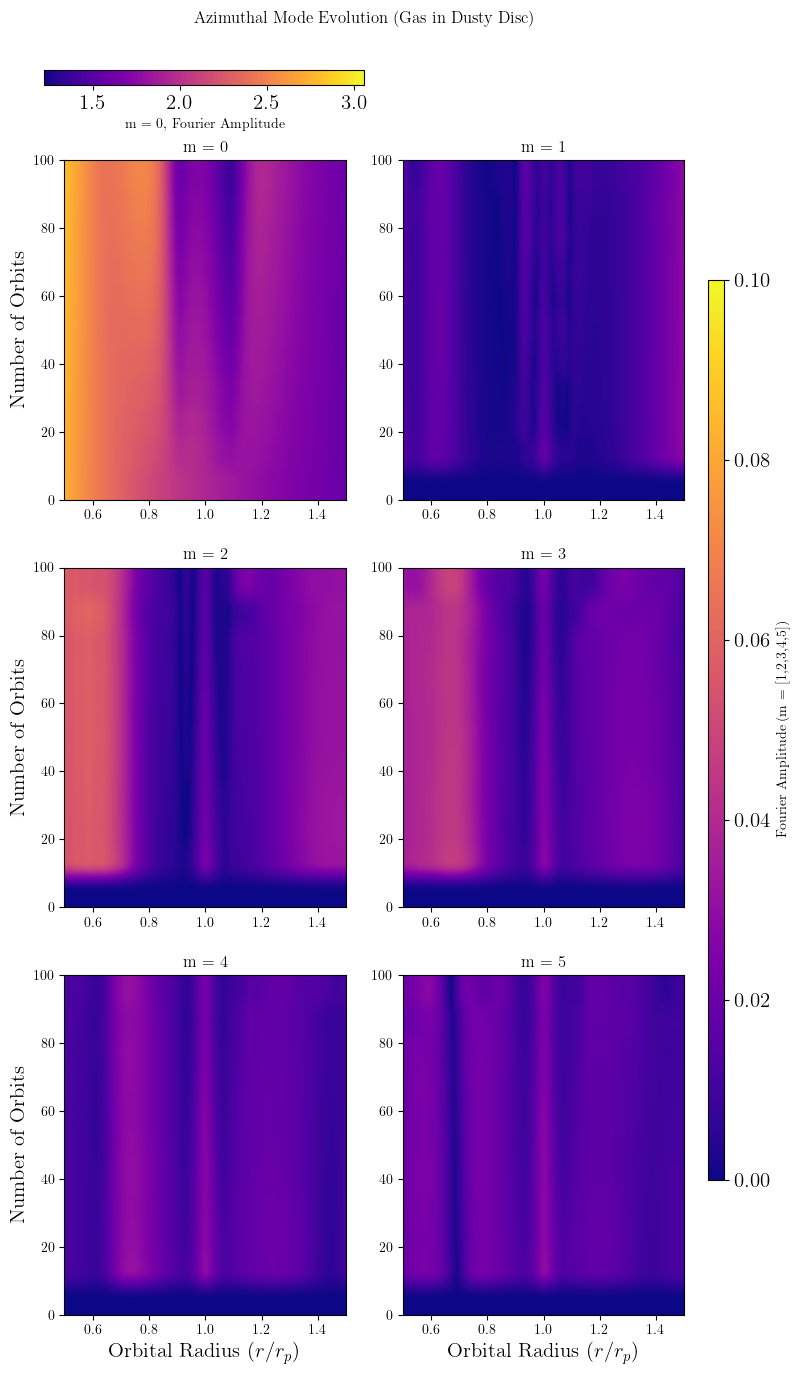}
\caption{Time evolution of the Fourier amplitudes of modes, $m = 0$ to $5$, for the gas in a Neptune planet embedded dusty disc with dust fraction $f_d = 0.01$, St = 0.2 and $\nu = 5 \times 10^{-7}$ . The evolution of the Fourier amplitudes for all modes show little variance in the region of the dusty ring location until $t = 80$ when the RWI is triggered, growing in amplitude from the outer gap edge and dusty ring location.} 
\label{fig:gas_in_dust_growth}
\end{figure}

Having shown that the addition of dust in a gaseous disc can cause the RWI to be triggered, we investigated the evolution of the perturbations through tracking the amplitudes of the Fourier modes in gas and dust density. 

In Fig. \ref{fig:dust_growth} we present the time evolution of the azimuthal Fourier components, $m = 0$ to $5$ for the disc previously in \ref{stokes} with dust fraction, $f_d = 0.01, \stokes = 0.2, \nu = 5 \times 10^{-7}$. We see that from modes, $m = 1$ to $5$, the Fourier amplitude at the location of the exterior dusty ring $(r/r_p = 1.2)$ grows from an early stage of the disc at a radius closer to the planet when it starts to form. These modes evolve together during the early stages and at around $t = 80$, modes $2, 3$ and $5$ grow faster than $1$ and $4$ which is around the time that the dusty RWI is triggered. Additionally we see the growth of the Fourier amplitudes around the planet $(r/r_p = 1.05)$ in $m = 1,2,3$, which would correspond to the asymmetric horseshoe region visible in Fig. \ref{fig:t100dens}. The large amplitude of the $m=5$ component reflects the number of vortices seen in Fig. \ref{fig:t100dens}.

When considering the gas in this dusty disc, Fig. \ref{fig:gas_in_dust_growth} presents the time evolution of the azimuthal Fourier components, $m = 0$ to $5$. We see that the evolution of the Fourier modes show little variance in the region exterior to the planets outer gap edge. However in modes 2 and 3 we can see limited growth in the region between the outer gap edge and dusty ring location from $t = 80$. 

\begin{figure}
    \centering
    \includegraphics[width=\linewidth]{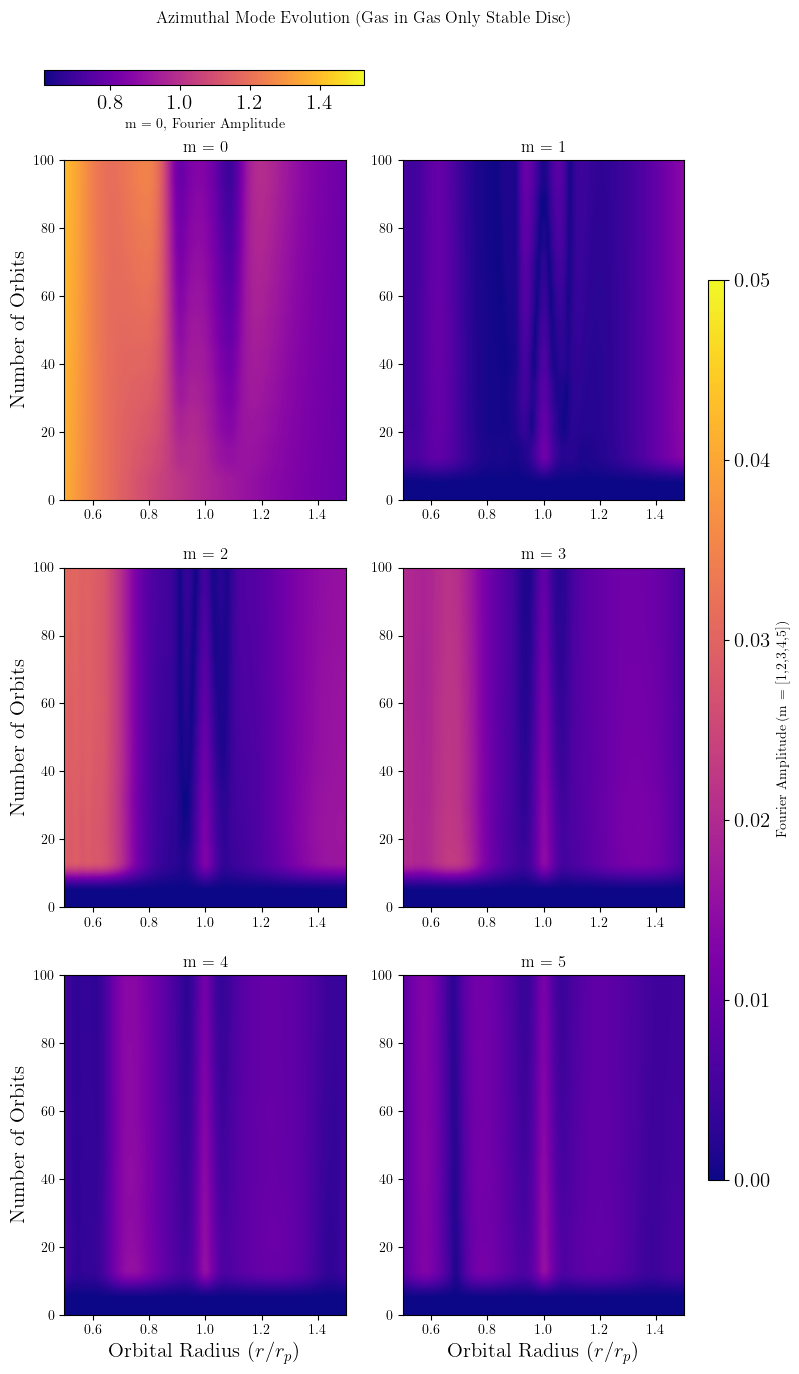}
\caption{Time evolution of the Fourier amplitudes of modes, $m = 0$ to $5$ for the gas in a Neptune planet embedded gaseous disc with $\nu = 1 \times 10^{-6}$. In this disc, the gas is stable up to $t = 100$, showing no features of instability growth in the outer pressure bump.} 
\label{fig:gas_in_stable_growth}
\end{figure}

When comparing to a purely gaseous disc in Fig. \ref{fig:gas_in_stable_growth} with the same kinematic viscosity we do not see these features where we know from Fig. \ref{fig:gas5e-7} that the RWI is not triggered in this gaseous disc by t = 100 orbits. This comparison and Fig. \ref{fig:dust_growth} shows that the addition of dust causes these non-axisymmetric features in the gas. 

\begin{figure}
    \centering
    \includegraphics[width=\linewidth]{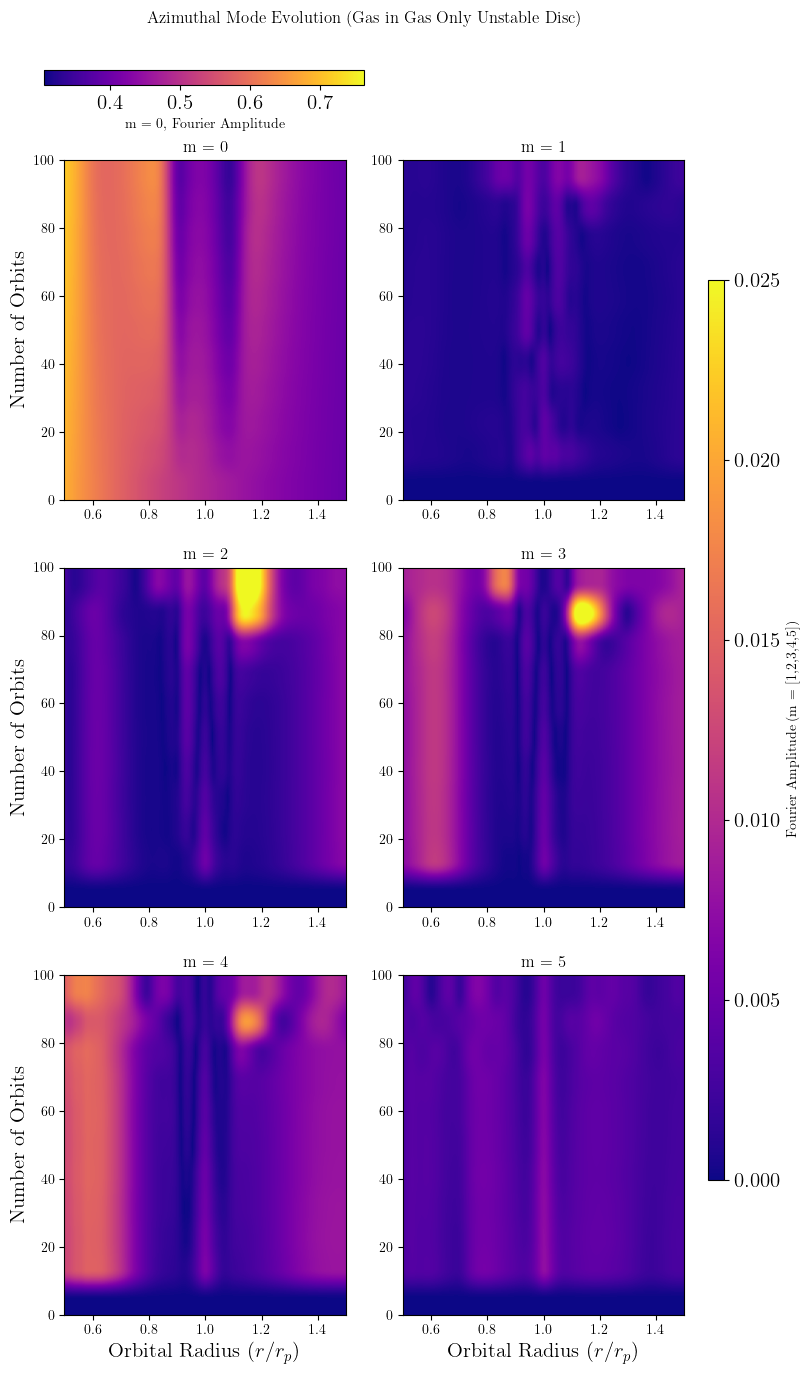}
\caption{Time evolution of the Fourier amplitudes of modes, $m = 0$ to $5$ for the gas in a Neptune planet embedded gaseous disc with $\nu = 1\times 10^{-7}$. For this disc, the gas triggers the RWI by t = 100 orbits and we see clear features in the plots for modes 2,3,4 indicating the growth of the instability.} 
\label{fig:gas_in_unstable_growth}
\end{figure}

Lastly we include a gaseous disc with $\nu = 1 \times 10^{-7}$ which does trigger the RWI in Fig \ref{fig:gas_in_unstable_growth} to show the features that can be seen in the time evolution of the azimuthal Fourier components for an unstable disc. We see that the region at the exterior pressure bump shows dominant growth in the Fourier amplitudes for modes 2,3 and 4 and the region undergoes instability from $t = 80$. These features are similar in shape and location as in Fig. \ref{fig:gas_in_stable_growth}, while the amplitude is lower relative to the rest of the disc. 

It should be noted that although for the particular cases studied here, the growth rates are similar, the dusty RWI differs in character compared to the gas-only RWI. The most unstable azimuthal wavenumber involved appears to be higher, and in general the scales of the perturbations appear to be smaller in the dusty RWI. This may have implications for the stability of the resulting vortices \citep{lovascio22}. Moreover, the dusty RWI requires a dust ring to be setup in a pressure bump. If this bump is unstable to the gas RWI, it may be the case that such a dust ring will not be able to form, and therefore that the dusty RWI will not operate. Therefore, it is conceiveable that the dusty RWI will not replace the gas RWI as the instability mechanism for steep edges in discs, but rather expands the parameter space in which these edges are unstable. 

\subsubsection{Increasing Dust Fraction and Stokes Number}

\begin{figure}
\begin{subfigure}{\linewidth}
    \centering
    \includegraphics[width=\linewidth]{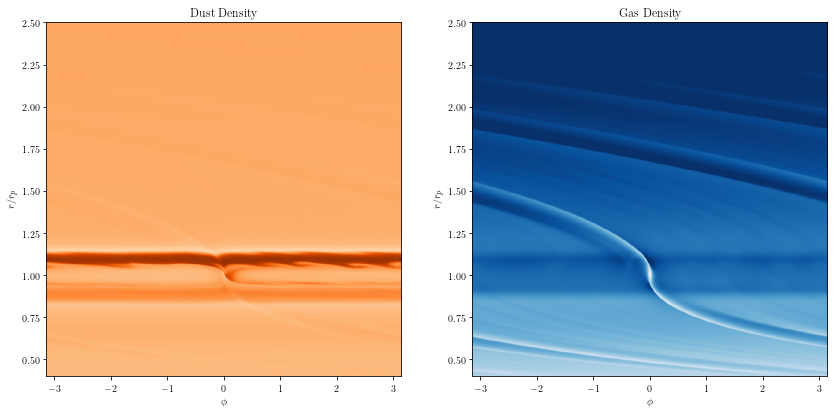}
\end{subfigure}
\begin{subfigure}{\linewidth}
    \centering
    \includegraphics[width=\linewidth]{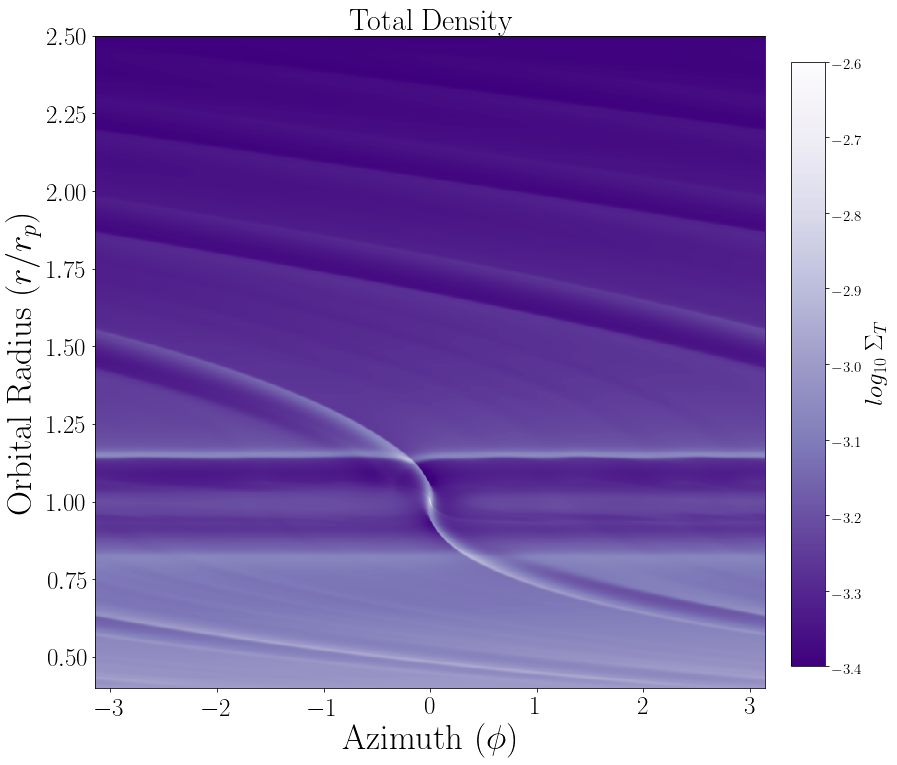}
\end{subfigure}
\caption{Density plots for dust (top left), gas (top right) and total (bottom) of the Neptune planet embedded disc with dust fraction, $f_{\rm d} = 0.1$ and Stokes number, $\stokes = 0.1$. The evolution of the disc is at $t = 50$ orbits. The dust ring formed exterior to the planet is smooth with vortices starting to be generated in the gap that the Neptune planet has carved for the dust phase which affects the gas phase and total density with weak spiral arms being launched through the disc.}
\label{fig:t50densmix}
\end{figure}

\begin{figure}
\begin{subfigure}{\linewidth}
    \centering
    \includegraphics[width=\linewidth]{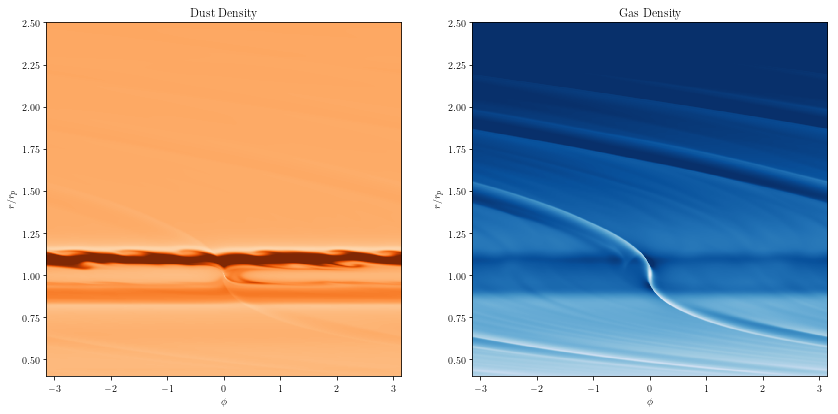}
\end{subfigure}
\begin{subfigure}{\linewidth}
    \centering
    \includegraphics[width=\linewidth]{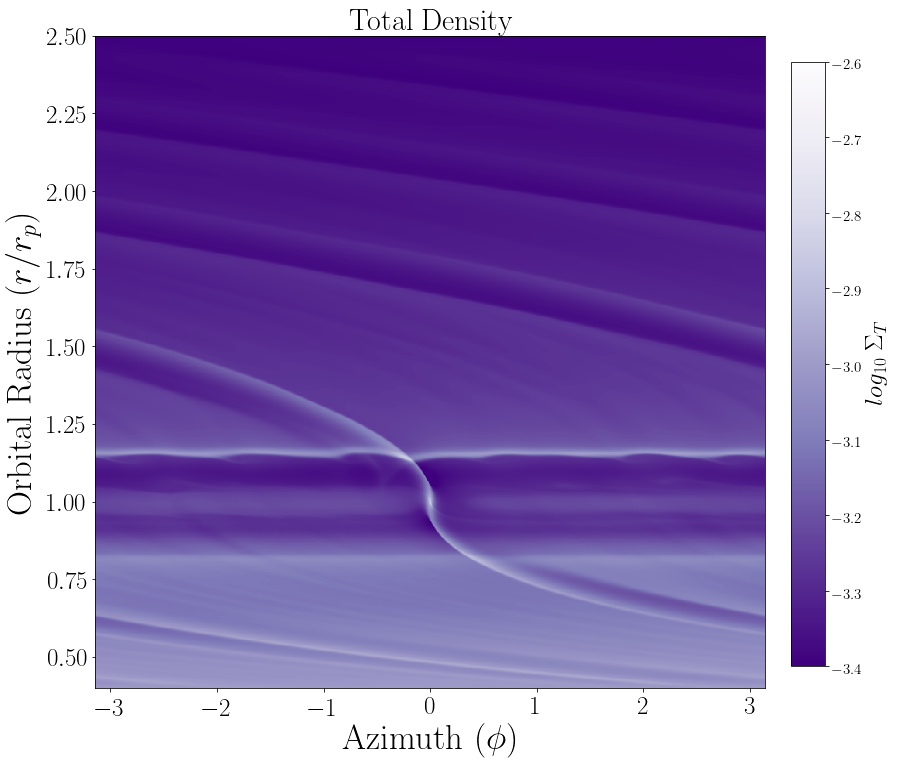}
\end{subfigure}
\caption{Density plots for dust (top left), gas (top right) and total (bottom) of the Neptune planet embedded disc with dust fraction, $f_{\rm d} = 0.1$ and Stokes number, $\stokes = 0.1$. The evolution of the disc is at $t = 60$ orbits. The dust ring has triggered the RWI at it's location after the gap edges became unstable closer to the planet. Compared to previous setup, this indicates addition of dust makes the disc even more prone to the RWI.}
\label{fig:t60densmix}
\end{figure}

With the previous case showing how the Stokes number of the dust can trigger the RWI at the ring location rather than at the gap edge, we wanted to test a case where both the dust fraction of the disc and Stokes number are increased to $0.1$, which is lower in Stokes number from the previous case but a much higher dust fraction than typical values for a disc. Expectations for this run was that the instability would trigger earlier as the dust could build up faster with the higher dust fraction.

In Fig. \ref{fig:t50densmix} we present the density plots for the dust, gas and total of the disc after 50 orbits. Looking at the total density plot we see that a dusty ring has been accumulated in the outer pressure bump as expected and the ring itself is smooth. Looking at just the dust plot however we see that interior to the dusty ring, the exterior gap that the Neptune planet has carved is unstable with vortices occurring already at each side of the gap edge with hints of the dust ring being distorted by the non-smooth gap edge. This is unexpected as the higher dust fraction case from subsubsection \ref{dustfracs} which has a dust fraction of $0.1$ but Stokes number $0.01$ does not have the same features in its early evolution. In the gas density plot we see that the vortices in the gap has accumulated enough material to launch weak spiral density waves through the disc. Overall however we see that the vortices in the gap for the dust plot do not affect the stability of the outer ring in the total density plots significantly as the combined gas and dust ring is generally smooth.

\begin{figure}
\begin{subfigure}{\linewidth}
    \centering
    \includegraphics[width=\linewidth]{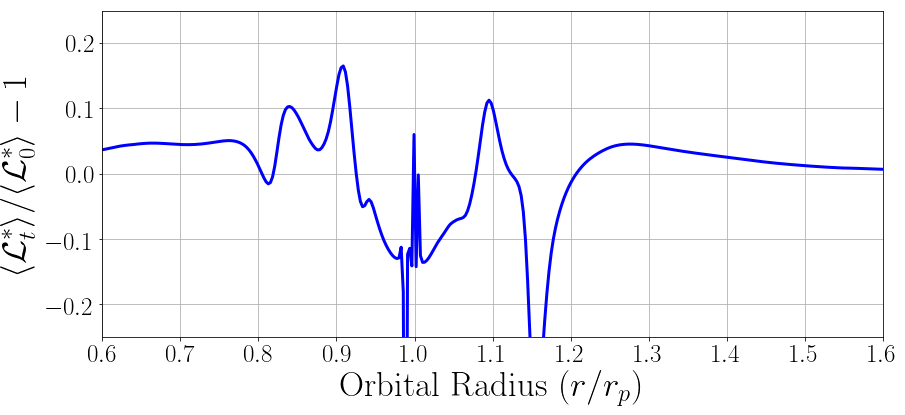}
\end{subfigure}
\begin{subfigure}{\linewidth}
    \centering
    \includegraphics[width=\linewidth]{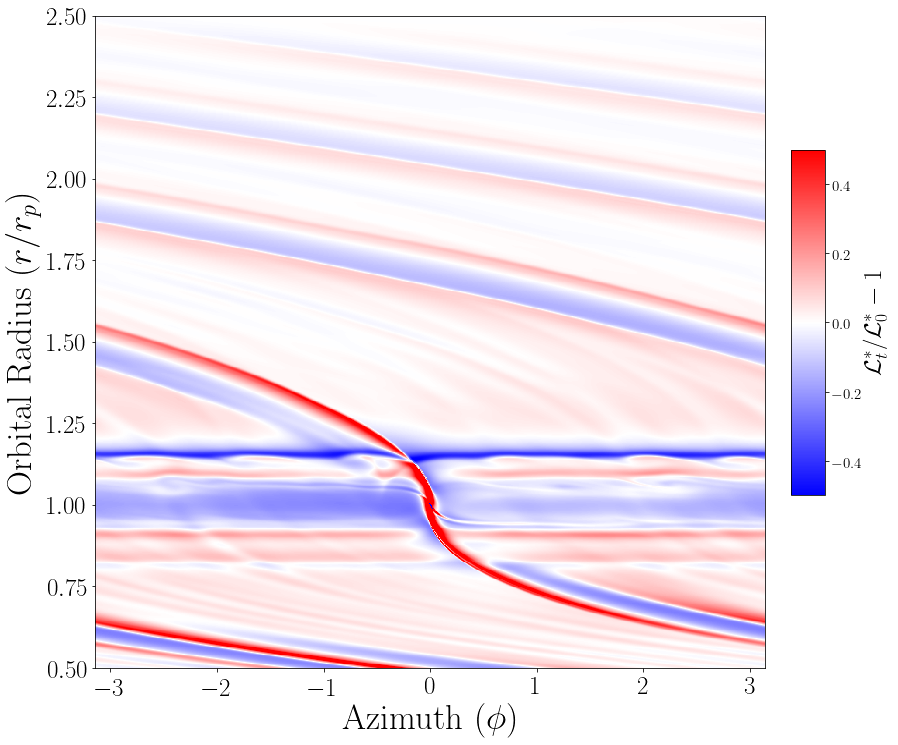}
\end{subfigure}
\caption{Radial profile of the azimuthally averaged "mixed" key function (top) and 2D key function plot (bottom) for a dust and gas disc with Neptune sized planet embedded at $1 r_{\rm p}$ after $t = 60$ orbits. The dust fraction is set to $f_{\rm d} = 0.1$, Stokes number, $\stokes = 0.1$ and kinematic viscosity, $\nu = 5\times 10^{-7}$. The radial profile shows an extremum at around $1.5 r_{\rm p}$ and the 2D plot shows the combination of the gap edge becoming unstable and the dusty ring triggering the RWI.}
\label{fig:highdensstkey60}
\end{figure}

Fig. \ref{fig:t60densmix} shows the evolution of the disc at 60 orbits and we see in the dust density plot that the dust ring has become unstable and the ring has triggered the RWI similar to the higher Stokes number case as we see the similar features in the ring of vortices being generated. This in turn has made the gas phase more unstable around the gap edge as we see small vortices in the gas density plots but are harder to discern. Overall the instability has triggered earlier than previous cases as we expected due to the faster build up of dust in the pressure bump exterior to the planet, however we do see that the gap has an effect on how clear the generated vortices are compared to the other cases. In this case, vortices generated at the outer gap edge dominate the early evolution before the dusty RWI vortices appear at the dusty ring location. This combination generates a "messier" evolution as the two locations compete, which makes it harder to discern the patterns.  Compared to the control case of the gas only disc we see that a combination of adding a higher dust fraction and higher Stokes number makes the ring exterior to the planet more unstable and a smooth outer dusty ring would be disrupted in a short timescale and more prone to instabilities when both parameters are increased. In this case the gap edge and ring has become unstable within 60 orbits, compared to the previous setup which trigger the RWI within 80 orbits. In terms of observation we would not see a clear smooth ring at this stage of its evolution in this case and the case of Stokes number of $0.2$ as the ring would be disrupted in a short timescale.

In Fig. \ref{fig:highdensstkey60} we show the radial profile and 2D plot of the "mixed" key function where we see in the radial profile a steeper extremum compared to subsubsection \ref{stokes} at the dusty ring's location. Compared to the previous setup of only increasing the Stokes number, we can clearly see that the vortices are not as well defined and the evolution of the outer region is more chaotic. Regardless, we do show that a higher dust fraction and Stokes number combined can make the regions exterior to the planet more prone to the RWI as the plots would indicate compared to the gas only case.

\subsection{Drag Cutoff} \label{cutoffsec}

The interpretation of the stability of dust rings in terms of the RWI neglects any possible effects of drag. In the thermodynamic interpretation of dusty gases (\citealt{lin17}), drag appears as a cooling term. Cooling is known to have an effect on the RWI itself in the case where cooling takes the system back to an equilibrium state (\citealt{huang22}). In our case, however, there is no equilibrium state as dust will continue to accumulate in the pressure bumps. Moreover, drag is known to create its own instabilities, such as the Streaming Instability (\citealt{youdin05}), and these are operating inside vortices as well (\citealt{surville19}, \citealt{lovascio22}). The simplest RWI interpretation of our results requires a separation of time scales: drag slowly sets up an RWI-unstable state, but does not play a role in the subsequent evolution of the instability. In order to test this idea, we ran simulations of the disc with the setup from section \ref{stokes} but with the drag modified. We ran the simulation of the disc up to the point before the RWI triggers at around 78 orbits, but then increased the collision rate between gas and dust by a factor of $10^9$. This in turn reduces the stopping time between the two phases by the same factor, and makes the two phases for all intents and purposes perfectly coupled. Therefore this limit eliminates the possibility of drag instabilities between the dust and the gas happening prior to the RWI triggering, since they require a finite stopping time. Note that the opposite limit of an infinite stopping time, while eliminating possible drag instabilities, is unfeasible since it violates the fluid approximation for the dust component (see e.g. \citealt{jacquet11}). 

\begin{figure}
\begin{subfigure}{\linewidth}
    \centering
    \includegraphics[width=\linewidth]{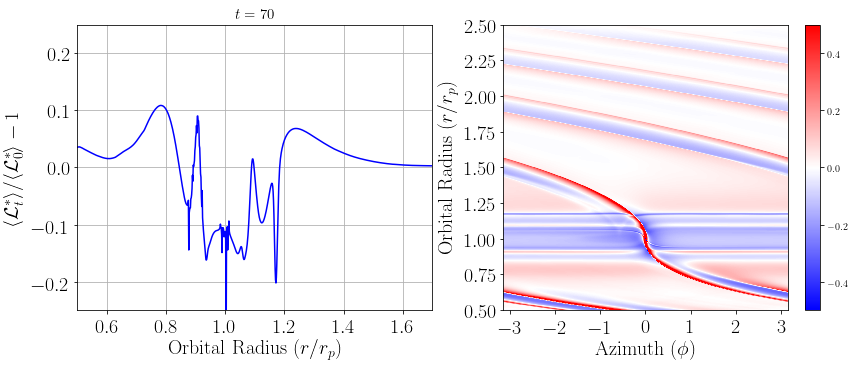}
\end{subfigure}
\begin{subfigure}{\linewidth}
    \centering
    \includegraphics[width=\linewidth]{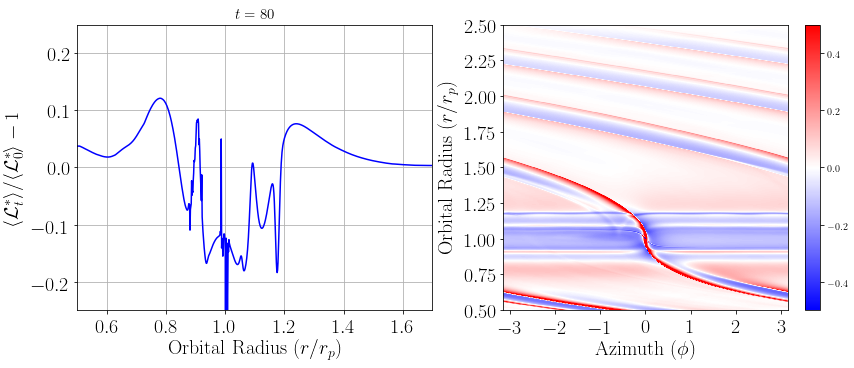}
\end{subfigure}
\begin{subfigure}{\linewidth}
    \centering
    \includegraphics[width=\linewidth]{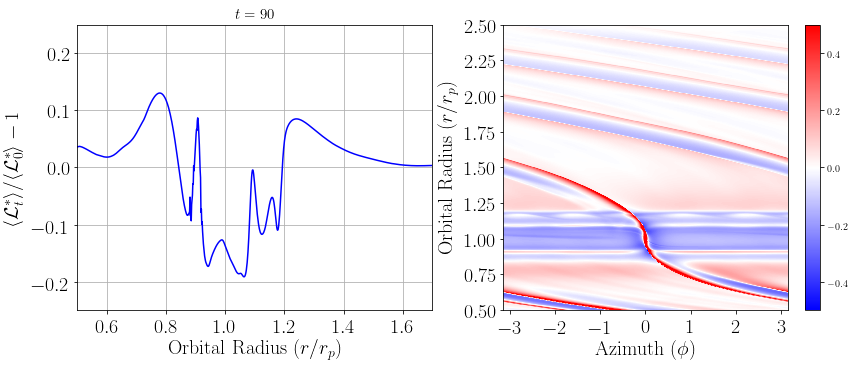}
\end{subfigure}
\caption{Time evolution of the "mixed" key function for a Neptune planet embedded with dust fraction, $f_{\rm d} = 0.01$ and Stokes number, $\stokes = 0.2$ at $t \in [70,80,90]$ orbits (top, middle, bottom). Perfect coupling is introduced at $t = 78$ through modifying the collision rate. We see that the RWI is responsible for the vortices generated at the dusty ring's location rather than drag instabilities.}
\label{fig:cutoffcombo}
\end{figure}

\begin{figure}
\includegraphics[width=\linewidth]{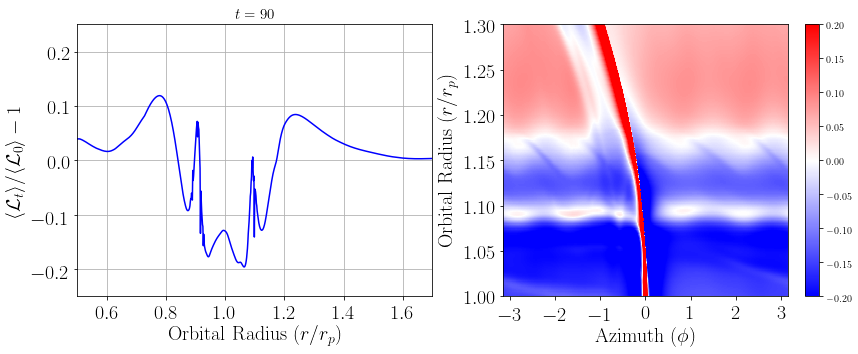}
\caption{Radial profile of the azimuthally averaged key function (left) and 2D key function plot (right) for the gas phase of the disc. The dust fraction is set to $f_{\rm d} = 0.01$, Stokes number, $\stokes = 0.2$ and kinematic viscosity, $\nu = 5\times 10^{-7}$. The 2D plot shows how the gas phase of the disc is affected by the dusty ring triggering the RWI through generation of vortices.} 
\label{fig:gascutoff}
\end{figure}

In Fig. \ref{fig:cutoffcombo} we present the evolution of the disc at times $t \in [70,80,90]$ orbits with a dust fraction, $f_{\rm d} = 0.01$ and Stokes number, $\stokes = 0.2$, with perfect coupling introduced at $t= 78$ orbits. From the radial profiles of the key functions we see that a minimum has formed at the dusty ring's location at $t = 70$ but has not triggered the RWI and generated vortices yet as shown in the 2D plot. When perfect coupling is introduced, the middle plots showing $t = 80$ the extremum starts to decrease as the RWI starts to trigger and finally in the bottom panel at $t =90$ we can see that the vortices have been generated by the RWI which acts to reduce the extremums that indicate the areas prone to the RWI. Since perfect coupling is enforced, the thin dust ring is not sustained as the RWI smooths out the density extremum. The emergence of vortices in the case of perfect coupling shows that it is the RWI triggering instead of drag instabilities at the location of the dusty ring. 

We clarify this further with Fig. \ref{fig:gascutoff} which shows the key function plots for the gas phase only. Since equation (\ref{keyfuncgas}) only includes quantities for the gas phase, in the radial profile we do not see the minimum caused by the dust ring. However, we can clearly see the effects of the dust ring triggering the RWI through the four vortices generated at the ring location that can be seen in the 2D plot of the key function for the gas phase. 

The radial profile "mixed" key function has been a valuable representation in indicating regions that are prone to the RWI in a dust and gas disc as the dust phase may trigger the RWI and affect the gas phase. As highlighted before and reinforced with the plots in this section, the width of the extremum seems to play a more important role than the amplitude in how prone a region is to the RWI. With the results showing clearly the generation of vortices in the absence of drag, we can conclude that the instability present during the evolution of the dusty ring in our simulations would be the RWI.

\subsection{Nonlinear phase of the Dusty RWI} \label{longvort}

In the previous sections, we have focused on the onset of the dusty RWI and the conditions under which it is triggered. Now, we study the evolution in the nonlinear phase where vortices have formed. 

\subsubsection{Formation and Structure of Vortices near the Dusty Ring} \label{initcomp}

For the onset of the RWI, a typical condition for the linear evolution of the instability (\citealt{lovelace99}; \citealt{li00a}) is the growth of the non-axisymmetric vortensity perturbations which in turn can form multiple vortices. After the linear regime, the RWI can enter the non-linear stage whereby the vortices form and merge into a larger vortex which impacts the disc dynamics and morphology significantly. Most recently, \cite{cimerman23} studied the evolution of the gas only RWI and the stability of the gap edges induced by a planet. They present a typical numerical evolution of the gas only RWI and a semi-analytical analysis of the instability. 

\begin{figure}
\includegraphics[width=\linewidth]{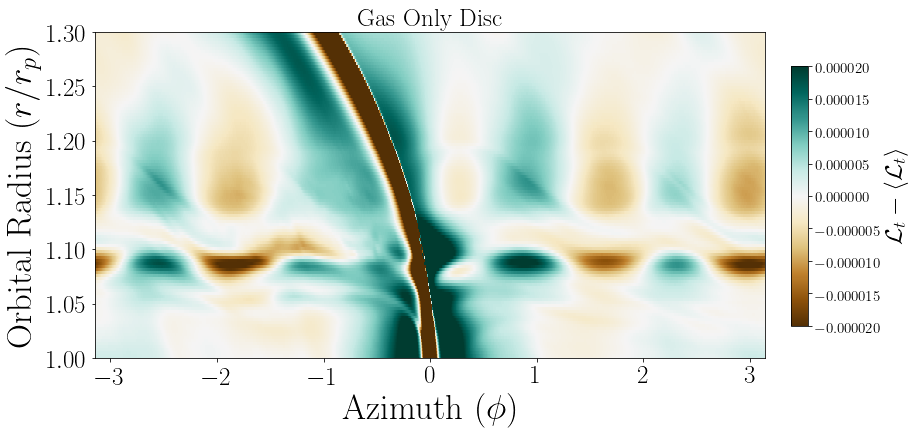}
\caption{Typical structure of linear evolution of gas only RWI. 2D plot of the non-axisymetric part of the gas only key function at $t =96$ orbits with kinematic viscosity, $\nu = 2.5\times 10^{-7}$. We see the formation of vortices exterior to the outer gap edge at $r = 1.15 r_{\rm p}$.} 
\label{fig:gasonlyrwi}
\end{figure}

We first present our gas only disc that triggers the RWI in Fig. \ref{fig:gasonlyrwi} to show the typical structure of vortices formed at the outer gap edge of a planet embedded disc. The plot shows the non-axisymmetric part of the gas only key function. The kinematic viscosity of this disc was set to $\nu = 2.5\times10^{-7}$ which is lower than previous setups to ensure the onset of the instability within 100 orbits. We see that four vortices have formed within 96 orbits and the inner gap edge is undergoing the RWI as well. These vortices will later merge in the non-linear stage and form two large vortices at the outer gap edge. This typical structure is similar to the studies by \cite{cimerman23} (see their Figs. 2 and 4), and we can use this as a comparison to what happens when we add dust.

\begin{figure}
\begin{subfigure}{\linewidth}
    \centering
    \includegraphics[width=\linewidth]{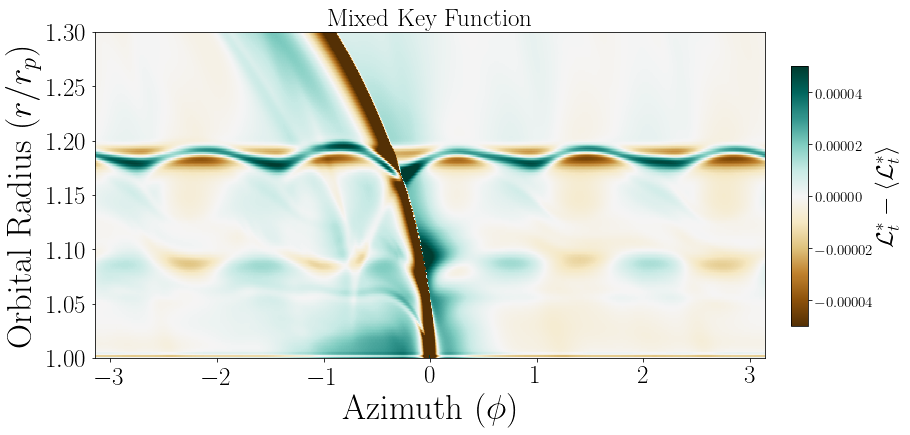}
\end{subfigure}
\begin{subfigure}{\linewidth}
    \centering
    \includegraphics[width=\linewidth]{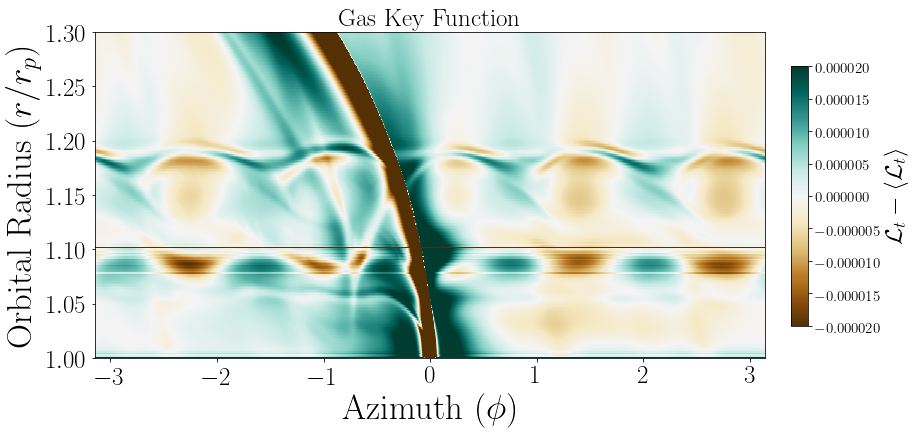}
\end{subfigure}
\caption{2D plot of the non-axisymmetric part of the mixed (top) and gas (bottom) key function at $t =100$ orbits. The dust fraction is set to $f_{\rm d} = 0.01$, Stokes number, $\stokes = 0.2$ and kinematic viscosity, $\nu = 5\times 10^{-7}$. We see the formation of vortices similar to the linear evolution of the gas only RWI at the outer gap edge with additional structures at the location of the ring.}
\label{fig:vorticest100}
\end{figure}

In Fig. \ref{fig:vorticest100}, we present the non-axisymmetric part of the mixed (top) and gas (bottom) key function for the dusty RWI in our previous disc setup of dust fraction, $f_{\rm d} = 0.01$, Stokes number, $\stokes = 0.2$ and kinematic viscosity, $\nu = 5\times 10^{-7}$. From the mixed key function plot we see that the addition of the dusty ring at $r=1.19~r_{\rm p}$ has created a region at its location which does not appear in the gas only RWI in the previous figure. When looking at the gas phase of the dusty disc, the gas key function plot at the bottom of Fig. \ref{fig:vorticest100} shows the typical structure of the onset of gas only RWI up to $r = 1.17 r_{\rm p}$ with the four vortices appearing at $r = 1.15 r_{\rm p}$. These however are overshadowed in amplitude by the structures at the dusty ring's location at $r = 1.19 r_{\rm p}$ which show stronger vortensity perturbations. The extra layer of vortices generated at the dusty ring's location dominates the evolution of the disc compared to the vortices generated at the outer gap edge. Note that for this level of viscosity, no vortices are generated within 100 orbits in a gas-only disc setup, previously seen in Fig. \ref{fig:gas5e-7}. This reinforces that the dusty ring is a cause for the triggering of the RWI at a separate location beyond the gap edge. 

The number of vortices that form initially is similar to the gas-only RWI (see e.g. \citealt{li00b}, \citealt{cimerman23}). We investigate the long term evolution of the dusty disc and the accumulation of material in these vortices in the next section. However, we first present the structure of the vortices where we analyse the interior of the vortices generated in our results at $t = 1500$ orbits. From the initial formation of these vortices at $t = 100$, material starts to accumulate into the vortex centres within $t = 200$ and the vortices begin to merge together, but the structure of the vortices evolving from $t = 200$ to $t = 1500$ remain very similar in configuration and size. A possible reason for this, shown previously, are that vortices tend to break up when loaded with dust (see \citealt{fu14b}, \citealt{lovascio22}) which would explain why the evolution of the size of the vortices remain similar as there would be a limit to how much dust a vortex can handle before it starts tearing apart and off-loading material to the surrounding area. This consistent size is explained and analysed more in the long term evolution section with density plots, but first we analyse the structure of the vortices.

\begin{figure}
\begin{subfigure}{\linewidth}
    \centering
    \includegraphics[width=\linewidth]{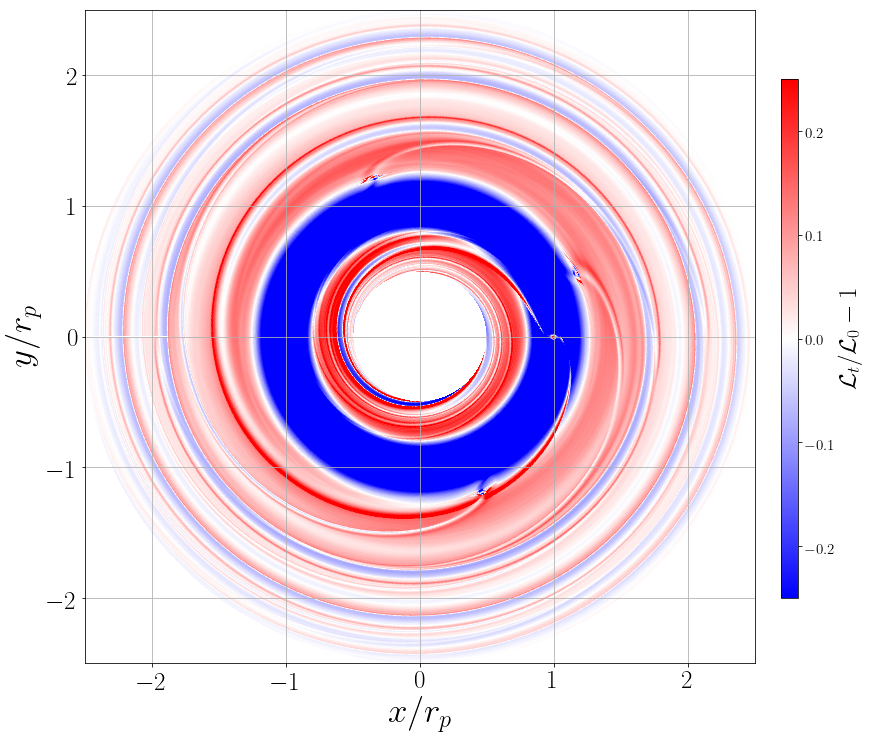}
\end{subfigure}
\begin{subfigure}{\linewidth}
    \centering
    \includegraphics[width=\linewidth]{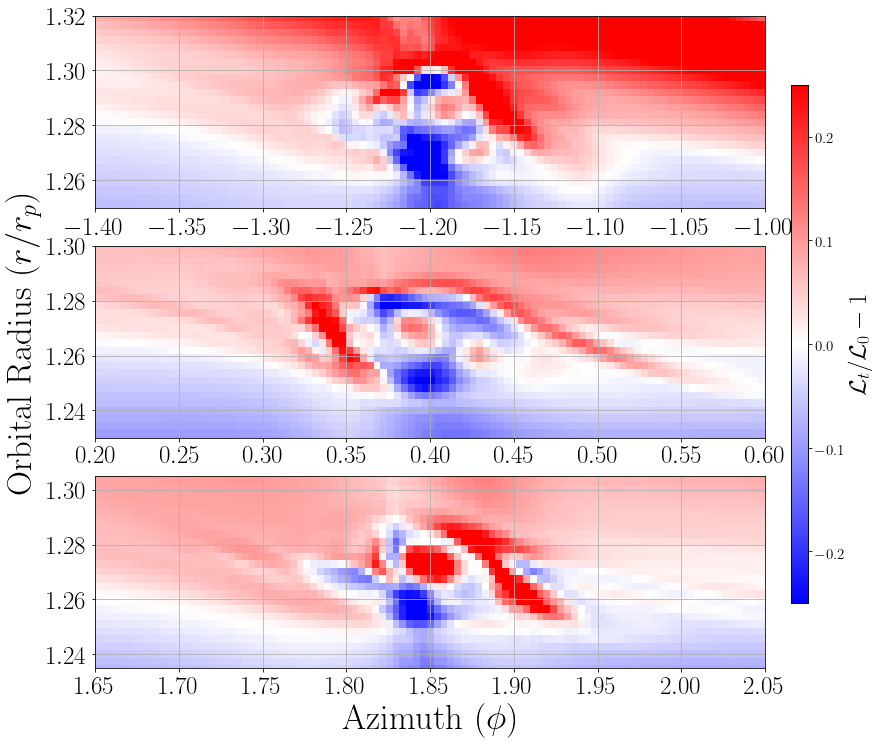}
\end{subfigure}
\caption{Polar plot of the "mixed" key function (top) and zoom in of the three vortices that have formed after $t = 1500$ orbits for for a Neptune planet embedded disc with dust fraction, $f_{\rm d} = 0.01$ and Stokes number, $St = 0.2$. We see that the centres of the vortices have accumulated a large amount of dust and the inner structure of the vortex is chaotic.} 
\label{fig:vortices_zoom}
\end{figure}

With Fig. \ref{fig:vortices_zoom}, we present the polar plot of the mixed key function in the top panel and a close up of the three vortices in the disc with the bottom three panels in the figure. We see that the three vortices have generated their own spiral density waves throughout the disc in the regions interior and exterior to the planet where these spiral density waves are weaker than the planet driven spiral density wave. At this snapshot we focus on two of the three vortices as one of them (first of the bottom three panels) is in the wake of the planet's spiral density wave, however we do note that the structure of this vortex is as complex as the other two. From the closeup of the vortices we first see that they do not have a uniform structure and the centres of the vortices are complex in terms of its key function which represents the density and vorticity. The cores of the vortices remain highly dynamic over their lifetime, with dust loading possibly destabilizing the flow, keeping material flowing around the vortex centres and not settling the in middle.

\subsubsection{Long Term Evolution} \label{long}

From the previous results, we see that the addition of dust can make the region exterior to the planet more prone to the RWI either through the gap edges or the dusty ring accumulated in the outer pressure bump. In this section we present the longer term evolution of the dusty ring after the RWI has triggered and vortices have been generated. The idea is that the planet-disc interactions create a pressure bump which accumulates the dust into a thin ring but the vortices generated from the dusty ring triggering the RWI would smooth out the density extremum. In this subsection we evolve the disc to 500 orbits in the first figure and consider the evolution of the vortices generated, the dusty ring and future implications for the disc as a whole with the second figure with evolution of the disc up to 1500 orbits.

\begin{figure}
\includegraphics[width=\linewidth]{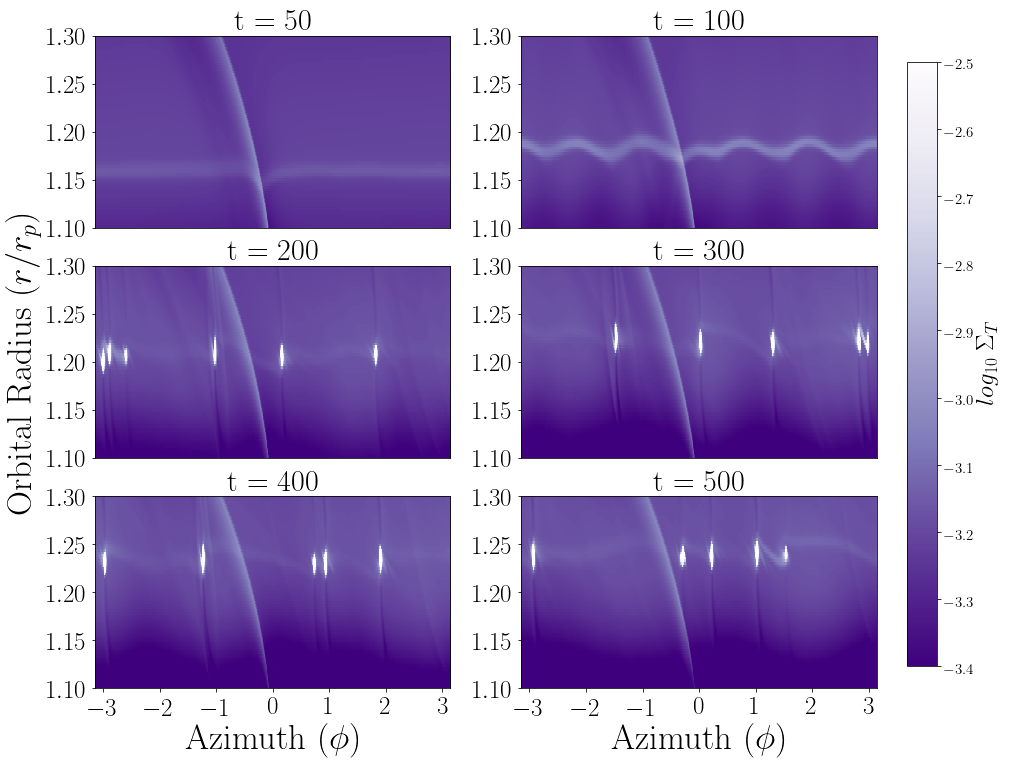}
\caption{Total density plots for region exterior to a Neptune planet that is embedded in a dust and gas disc with dust fraction, $f_{\rm d} = 0.01$, Stokes number, $\stokes = 0.2$ and kinematic viscosity, $\nu = 5\times10^-7$. The plots shown are the time evolution of the disc, $t \in [50,100,200,300,400,500]$.} 
\label{fig:long_mig}
\end{figure}

In Fig. \ref{fig:long_mig} we present the evolution of the total density of the outer region of the planet up to 500 orbits with the previous setup of $f_{\rm d} = 0.01$, $\stokes = 0.2$ and $\nu = 5\times10^{-7}$. As previously seen up to 100 orbits, the dusty ring that is accumulated in the pressure bump triggers the RWI and the smooth ring is distorted around the disc. The vortices at the dusty ring's location start to accumulate material from the ring and the surroundings towards the centre of the vortex which is highlighted further in the middle left plot of $t = 200$ where vortices can be seen to have a considerable amount of material accumulated and the dusty ring is now fainter. From the next evolution plot of $t = 300$, two of the vortices have merged into one, where we can see now five vortices in the outer region of the planet. Compared to a gas-only RWI, the vortices appear much stronger in density because of dust accumulation. Another difference is that the dusty vortices are continuously destroyed and regenerated. This is different to typical unstable gas-only disc cases (e.g. rightmost panel in Fig. \ref{fig:ngasonlycombo}) where the vortices tend to merge into a larger vortex within an earlier timescale. From these initial plots we see that the vortices generated by the dusty RWI exterior to a planet could in principle be locations of planetesimal formation as they are able to accumulate material which radially averaged density values exceeds unity of the dust to gas ratio. In the radial profiles which are not presented we see that the vortices accumulate dust close to 3-4 times the gas density at that radial location after the 500 orbits. Observationally, the dust ring itself will at this point be much harder to see as the vortices accumulate all the material. 

\begin{figure}
\begin{subfigure}{\linewidth}
    \centering
    \includegraphics[width=\linewidth]{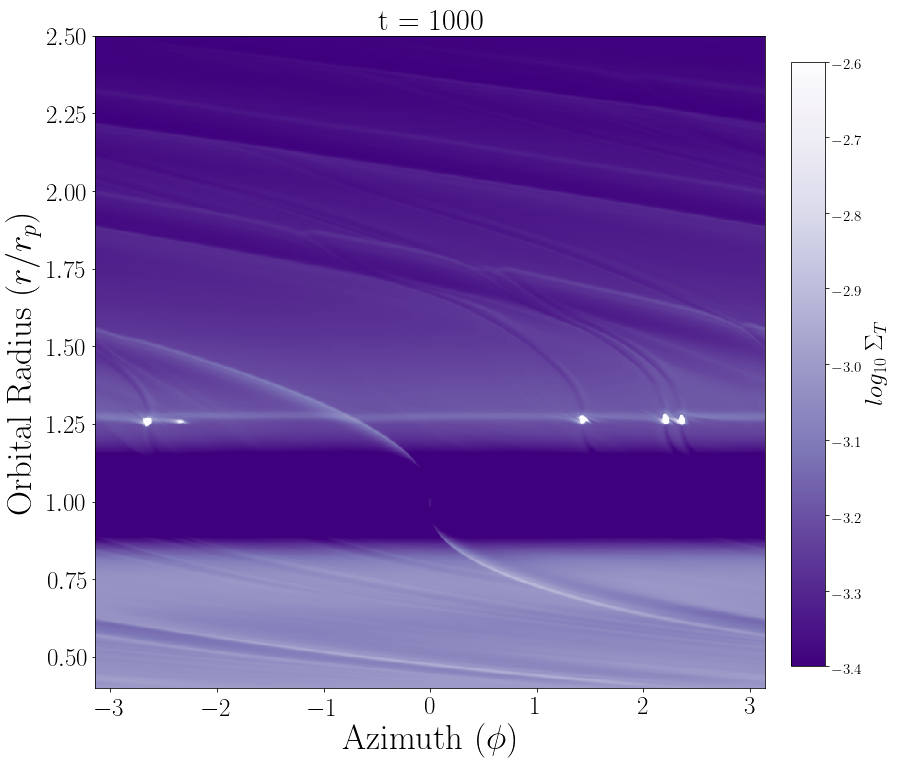}
\end{subfigure}
\begin{subfigure}{\linewidth}
    \centering
    \includegraphics[width=\linewidth]{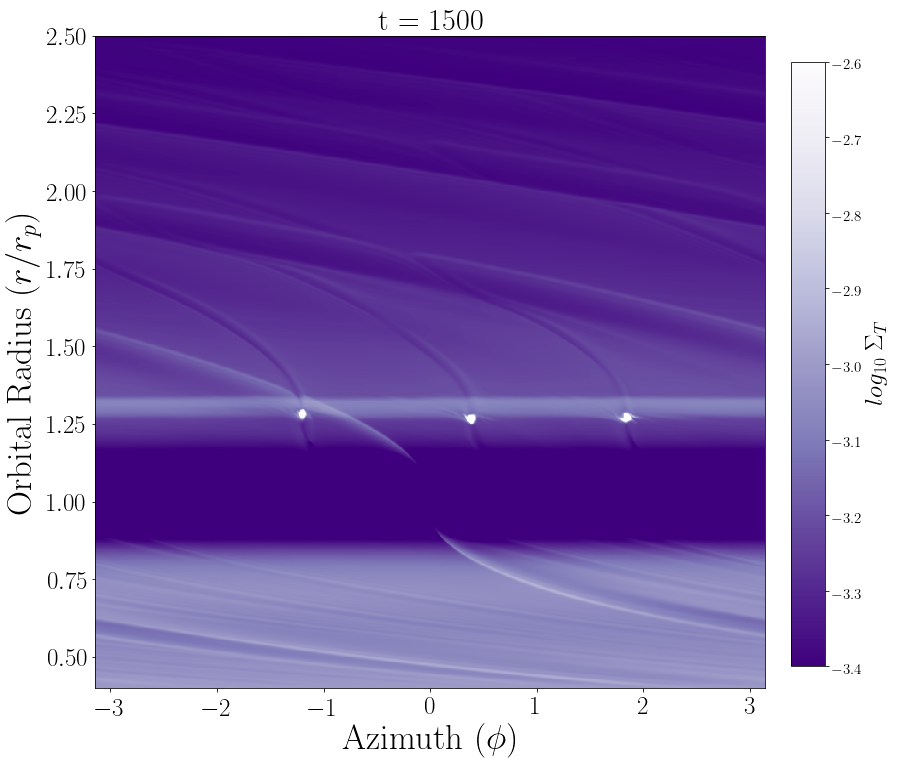}
\end{subfigure}
\caption{Total density plots for the disc at $t = 1000$ (top) and $t = 1500$ (bottom) with dust fraction, $f_{\rm d} = 0.01$, Stokes number, $\stokes = 0.2$ and kinematic viscosity, $\nu = 5\times10^-7$. A secondary wider dusty ring is shown to be forming exterior to the locations of the vortices which have merged into three distinct points.} 
\label{fig:long_times}
\end{figure}

An important feature shown in the evolution of the disc is that the dusty ring which started at around $r = 1.16 r_{\rm p}$ is distorted by the generated vortices and moved outwards from $t = 50$ to $t = 100$. Further evolution shows that the vortices which have accumulated the material start to migrate outwards as after 500 orbits, the radial location of the vortices generated and the material accumulated reach $r = 1.24 r_{\rm p}$. The migration of vortices has been studied before in \cite{paardekooper10} where the asymmetries in the spiral density waves launched the vortices can cause a net exchange of angular momentum which leads to the migration of the vortices. In the study of the gas only case, vortices move to regions of higher gas pressure and in our results we see that the dusty vortices migrate towards the pressure maxima in the outer pressure bump which is shifted outwards during the evolution of the disc, which occurs in our gas only disc as well with the planet carving a gap. Therefore, vortices as well as the remnants of the dust ring end up at the pressure maximum, the latter through usual radial drift. 

Evolution of the disc up to 1500 orbits is presented in Fig. \ref{fig:long_times} where we present the density plots at $t = 1000, 1500$. Comparing the evolution between $t = 500$ from the previous figure and $1000$ we see that a dusty ring is accumulating exterior to the radial location of the vortices. We note that this is a different ring to the original dusty ring that has gathered into the vortices. This is more clearly seen after another 500 orbits in the lower plot of the figure where a wide dusty ring is present and the vortices have merged into three distinct points. During the evolution of the disc from the formation of the vortices up to around $t = 1000$, the vortices act as a barrier to dust drifting inwards as it gets accumulated into the vortex centres. From $t = 1000$ and onwards, the dust density in the vortices oscillate between two limits, which would indicate that the vortex is going through an instability, causing it to lose material, due to the loading of dust. We find that the radial profile of the outer pressure bump does not have a "peak" anymore compared to the gas only simulation but instead the region exterior to the vortices has a flat profile. This would explain the observed morphology of the disc as the material drifting inwards is building up into a flat pressure trap, generating a wider ring compared to the thin dusty ring first observed before the RWI triggered. 

In previous studies of gas only RWI (e.g. \citealt{lin12}; \citealt{meheut12a}; \citealt{rometsch21}), the long term evolution results in the merging of vortices into a larger, more elongated vortex that is stable depending on the strength of the decaying mechanisms such as the elliptical instability \citep{lesur09}. In the study of dust loaded vortices (e.g. \citealt{surville19}; \citealt{lovascio22}), the presence of dust severely limits the survivability of large vortices (radial size comparable to a scale height) through a drag instability. In our simulations, we observe that vortices break up again after merging, presumably as a result of this instability. These dusty vortices remain over a long evolution and indicates that they are regulated by the amount of dust in the vortex cores as they go through phases of tearing and reformation from the surrounding material in the dusty ring. This has important implications in observing dust loaded Rossby vortices as in the long term they do not grow into the azimuthally extended structures observed in gas-only simulations. On the other hand, it is possible that dust, even when not dynamically important, will sink to the core of the vortex so that even in that case the dust distribution will be very compact in azimuth. These vortices however are long lived in our simulations of up to 10,000 orbits which suggest that in the future with higher resolution imaging of protoplanetary discs, these dusty Rossby vortices may be found in low viscosity regions.

Overall these simulations and results act as a possible cause for further study in the long term as we show that the addition of dust can cause the gas phase to become unstable to the RWI. This in turn could trigger locations of planetesimal formation and as shown, phases of observable rings in the disc. 

\section{Summary and Discussion} \label{discuss}

In this paper we set out to investigate the stability of dusty rings which have been formed in the outer pressure bump generated by planet-disc interactions with a Neptune sized planet. We studied specifically rings which triggered the RWI to explore the implications of its evolution in terms of possible planetesimal formation locations and observable features.

\subsection{Effects of Dust on the Proneness to the RWI}

From \cite{lovelace99}, the RWI can be triggered by a narrow ringed gas structure. Further studies as mentioned before (e.g. \citealt{meheut12a}; \citealt{cimerman23}; \citealt{chang23}) focuses on gas structures and the implications of the gas triggering the RWI, generating the vortices in the gas phase which then can affect the dust distribution in a disc. In this study we have found that the dust phase should also be a focus in a pressure bump when considering conditions for the RWI to trigger. The narrow dust ring structure which is observed in the results can alter the evolution of the region through the RWI which would be stable in gas only simulations. With the addition of higher dust fraction, initial results show that the outer pressure bump region can be more prone to the instability, triggering at an earlier timescale. We showed that this can occur in lower than expected viscosity discs however it would be interesting to see if the dust phase could trigger the RWI in higher viscosity discs which would not be able to without the additional substructures through similar narrow rings, for example with an even higher Stokes number, before the viscosity dampens the conditions for the instability to trigger. We point out that although the RWI has been observed and studied in simulations, there is no analytic criterion for the RWI. Efforts have been made to constrain the RWI (\citealt{chang23}), however the exact nature of the RWI needs further study in terms of the gas and dust phases.


We have shown in the results that if the RWI is triggered by the gap edges or the dusty ring itself, within a few hundred orbits, the original ring structure is accumulated into the vortices generated by the instability. In terms of observations, narrow ring like structures in low viscosity regions of a disc would not survive and be very short lived. The vortices generated exterior to a reasonably sized planet however have been shown to last thousands of orbits in this study. The issue for observing these with current advances in observations is the small scales of the vortices generated. In this study they grow radially to an extent of less than a scale height. The observation of small scale structures in a protoplanetary disc have been limited and the origin of the small scale structures observed is unclear. For example in \cite{tsukagoshi19}, they report an AU-scale dust clump which they hypothesise to be material accumulating in a vortex or a massive circumplanetary disc around a Neptune sized planet embedded in a disc. In the future we hope advances would lead to better resolved small scale substructures to study their origins.

The size of the dusty vortices generated reach up to a scale height in the radial direction past the linear evolution of the RWI and are long lived through regulation of the amount of dust material accumulated in the centres. The stability of these vortices are maintained through the onset of the RWI triggered at the dusty rings' locations against the drag instabilities tearing the vortex apart through loading of dust. Observations of these dusty Rossby vortices in low viscosity regions would require higher resolution imaging of protoplanetary discs.

\subsection{Long Term Evolution Implications}

Although the original ring structure in our study triggers the RWI and generates vortices that deplete the material from the ring, the long term evolution of the disc shows possible formation of a ring firstly at the radial location of the vortices and then spreading out to a wider ring that sits exterior to the vortices. The result of the stages of the evolution from $t = 100$ to $t = 1500$ show a wide range of future evolutions of the disc especially due to caveats in our simulation which would play important roles at different stages. Firstly with the formation of vortices and the accumulation of material into the centres, clumping of material and formation of planetesimals can aid the formation of planetary cores. With the dust buildup in a wider radial extent seen in later times, it provides a region that is more prone to the Streaming Instability which would further develop the formation of dust clumps. 

The radial migration of the dust ring and the vortices can have important consequences for the interpretation of observations, as the relative location of the (possibly observable) ring and vortices and the (unseen) planet changes with time. We saw that after 1500 orbits, the dust ring has increased its radial distance to the planetary orbit by a factor of 2.   
\subsection{Limitations and Outlooks}

In our simulations, we held the Neptune sized planet on a fixed orbit around the central mass as we wanted to explore how the planet-disc interactions would affect the onset of the RWI in the outer pressure bump. For a more accurate evolution of the disc the migration of the planet would be important as previous studies (e.g. \citealt{fu14}) showed that isolated vortices would decay on viscous timescales. Expectations for this would then be vortices that do not survive as long as our results if the planet migrates away in addition to the shown migration of vortices to an outer radial location. The onset of the RWI would be inhibited as well since a narrow dust ring would have to be formed to trigger the instability in the way shown in our results. 

Additionally in this study, we do not include self-gravity which would be important to the evolution of areas where dust is being accumulated such as the centres of the vortices and the dusty ring. The inclusion of self-gravity would aid in bringing more material into these regions that could then be used to grow grain sizes. In \cite{lovascio22}, they find that larger dust grains in vortices causes a drag instability to occur sooner which tears the vortex apart. 

In the results for the long term evolution, we utilise a single value for the Stokes number. For a more accurate evolution of the disc, multiple values for the stopping time would need to be considered. In \cite{mcnally21}, they study the growth rates of the Streaming Instability in a continuous range of dust sizes versus single sized where the fundamentals of the Polydisperse Streaming Instability (PSI) are described in \cite{paardekooper20}. In many cases with an interstellar power law distribution of dust sizes they found that the PSI grew more slowly than in single dust size cases. The possible implications for this as discussed in \cite{lovascio22} is that vortices could be more stable. Additionally they found that an enhancement of large particles increased the growth rate of linear PSI as long as the dust to gas ratio was above unity. With the long term evolution plots showing regions where dust grains could grow with a single dust size and dust to gas ratios above one, further research into the possibly of the same features created with multiple dust sizes would be interesting as the outcomes of the vortices generated by the RWI would be an important factor in the discs future evolution. 

We study the RWI and the resulting vortices in 2D. However when simulating vortices in 3D, they are shown to be unstable to the elliptic instability (\citealt{lithwick09}; \citealt{lesur09}; \citealt{railton14}) with \cite{richard31} showing how the aspect ratio of RWI vortices affects it's survival rate against the elliptic instability. \cite{lin18} showed that self gravity aids the growth and survival of 3D vortices through it's early stages where the elliptic instability can destroy it in a short timescale.

Other factors that would affect our study would be the planet mass and dust diffusion. Both play an important role in the formation, survival and size of the dust ring where we showed could trigger the RWI. Since dust diffusion would create a higher width ring, expectations would be regions which are less prone as we showed a narrow dusty ring is more prone to the RWI. For different planet masses, the size and shape of the pressure bump exterior to the planet would affect the build-up of dust. Further study will be needed on the relation between planet masses and the dusty RWI. 

Lastly, we consider a locally isothermal disc. Without the assumption of instant heating/cooling, gas thermodynamics have been shown to have a great affect on the survival and strength of vortices (\citealt{pierens18}; \citealt{tarczay20}), and, in addition, cooling has an effect on the RWI itself (\citealt{huang22}). Study of the dusty RWI and vortices generated in a non-isothermal disc could therefore lead to a different long term evolution of the disc.

With the inclusion of any of the factors above, the essential question in future works should be how long do the vortices survive for and how useful they are in aiding planet formation. One interest would be the possibility of "inside-out" planet formation (\citealt{chatterjee14}) whereby a gravitationally unstable ring of material would induce planet formation via core accretion and create a new pressure bump that gathers material drifting inwards, therefore repeating the process.

\section{Conclusion} \label{conc}

We performed 2D hydrodynamical simulations of a global disc to investigate the effect of dust on the onset of the Rossby Wave Instability in the pressure bump region generated by an embedded Neptune sized planet. We compared a gas only disc to a dust and gaseous disc and found that a narrow dust ring structure in the pressure bump, can trigger the RWI and make the region more prone to the RWI compared to a gaseous disc. This was through Stokes number, $\stokes = 0.2$ for the dust and typical dust fraction of $1\%$. The size of the dust grains was shown to be a more important factor in how prone the region was compared to initial dust fraction. 

In terms of stability of the dusty ring, the original structure was distorted and accumulated into the vortices generated by the RWI. This provides very little in terms of observations due to how short lived the dusty ring is. However, when considering long term evolution, we show a result whereby the vortices generated by the RWI can migrate outwards and in turn stall material drifting inwards generating a secondary dusty ring after a thousand orbits. This second ring survives for much longer for observations and this could lead to many potential different evolution that can aid formation of planetary cores through other physical effects such as the Streaming Instability.

\section*{Acknowledgements}

We thank Francesco Lovascio for useful discussions that contributed to this work. This work was performed using the DiRAC Data Intensive service at Leicester, operated by the University of Leicester IT Services, which forms part of the STFC DiRAC HPC Facility (www.dirac.ac.uk). The equipment was funded by BEIS capital funding via STFC capital grants ST/K000373/1 and ST/R002363/1 and STFC DiRAC Operations grant ST/R001014/1. DiRAC is part of the National e-Infrastructure. KC is funded by an STFC studentship. 

\section*{Data Availability}

FARGO3D is publicly available at \url{https://bitbucket.org/fargo3d/public.git} and all simulation setups will be shared on request to the authors.


\bibliographystyle{mnras}
\bibliography{mnras_guide} 


\bsp	
\label{lastpage}
\end{document}